# Quasinormal Mode Expansion Method for Resonators with Partial-fraction Material Dispersion

Xianshun Ming

*State Key Laboratory of Transient Optics and Photonics, Xi'an Institute of Optics and Precision Mechanics (XIOPM), Chinese Academy of Sciences (CAS), Xi'an 710119, China*

mingxianshun@opt.ac.cn

**Abstract**: In this paper, we first establish a Quasinormal Mode (QNM) solver for open resonators made of materials with general dispersion which can be modeled by partial fractions, and develop the corresponding analytical QNM expansion method (QNMEM) for both discrete and periodic resonant structures. When the response of the resonators is dominant by several leading QNMs, a simplified QNMEM can be used to analyze their spectra in a reasonable accuracy. The simplified QNMEM is used to analyze the spectra of the metal-dielectric-metal perfect absorber, which has the advantages of both high computation speed and clear physical insight.

The field enhancement and enhanced spectral response of nano/microresonators are connected with their resonant modes i.e., Quasi-normal modes (QNMs). Here we first establish the QNM expansion method (QNMEM) to rigorously compute the spectra of nonperiodic/periodic resonators. The basic idea of this method is to expand the scattered field into linear superposition of QNMs. Later, we will use the simplified QNMEM to analyze periodic subwavelength perfect absorber (PA) with only several leading QNMs retained.

The treatment of material dispersion is critical for QNMEM. When the material is dispersionless, the source-free Maxwell's equations are linear, thus it is quite straightforward to solve the QNMs from the eigen equations. Vial et. al. established the QNMEM for dispersionless material to treat scattering of aperiodic nanoresonators and diffraction of periodic resonators (resonant gratings) [1] , and later applied the simplified QNMEM to the design of mid-infrared PA [2] , where the material dispersion was taking into account by setting a linearization point iteratively. Absorption computed by the simplified QNMEM match well that by full wave FEM, but this iterative method only converges fast in the range where the dispersion of permittivity varies slowly, and calculate one QNM at a time. When the material constituting the resonator is dispersive, the source-free Maxwell's equations become nonlinear, making the problem more complex, one of the techniques to treat the nonlinearity is to introduce auxiliary field to linearize the original eigen equations, which can compute a series of eigenvalues "at one computation". For materials whose dispersion can be described by the Lorentz-Drude model, Yan et. al [3] established the QNMEM mainly for aperiodic nanoresonators by introducing auxiliary polarization **P** and current density **J**, and Gras et. al. [4] further developed the method for resonant grating with fixed incident angle.

The auxiliary fields introduced by Yan [3] and Gras [4] are for Lorentz-Drude model, which is best suitable to deal with resonators made of metal [5] or high doping semi- conductor [6]. For more general cases, the material dispersion can be modeled in a universal Partial Fraction model [7] . Here we will develop the auxiliary field method aiming at the Partial Fraction model, define new form of auxiliary fields to reformulate the linearized augmented eigen equations and solve the QNMs. Besides, the unconjugated form of the Lorentz reciprocity theorem and the Poynting theorem of the augmented Maxwell's equations, as well as the bi-orthogonality and normalization of QNMs are derived, also the semi-analytical form of excitation coefficients is obtained. Finally, we built a relatively general QNMEM for scattering of aperiodic nanoresonators and diffraction of resonant gratings, and use the simplified QNMEM to design PA.

1. **Definition and solving of QNM**

    1.1 **Concept of QNM and the related perfect absorbing mode**

    A QNM of open nanoresonators is the solution of source-free Maxwell's equations with outgoing wave (OWC) condition as follows:

$$\nabla \times \mathbf{E} = -\mu_0 \mu_r \frac{\partial \mathbf{H}}{\partial t},$$
$$\nabla \times \mathbf{H} = \epsilon_0 \epsilon_r \frac{\partial \mathbf{E}}{\partial t}. \qquad (1)$$

Assuming a time harmonic form of solution as $\tilde{\mathbf{E}}_m(\mathbf{r}, \widetilde{\omega}_m) = \tilde{\mathbf{E}}_m(\mathbf{r})\exp(-i\widetilde{\omega}_m t)$ and $\tilde{\mathbf{H}}_m(\mathbf{r}, \widetilde{\omega}_m) = \tilde{\mathbf{H}}_m(\mathbf{r})\exp(-i\widetilde{\omega}_m t)$ considering the OWC condition, Eq. (1) can be reformulated as

$$\begin{bmatrix} 0 & -i(\mu_0\mu_r(\widetilde{\omega}_m))^{-1}\nabla \times \\ i(\epsilon_0\epsilon_r(\widetilde{\omega}_m))^{-1}\nabla \times & 0 \end{bmatrix} \begin{bmatrix} \tilde{\mathbf{H}}_m(\mathbf{r}) \\ \tilde{\mathbf{E}}_m(\mathbf{r}) \end{bmatrix} = \widetilde{\omega}_m \begin{bmatrix} \tilde{\mathbf{H}}_m(\mathbf{r}) \\ \tilde{\mathbf{E}}_m(\mathbf{r}) \end{bmatrix}, \qquad (2)$$

where quantities related with the QNM are marked with "~". In general, Eq. (2) defines a non-Hermitian system, and the boundary condition is the natural boundary of the open space. The eigenfrequency $\widetilde{\omega}_m$ is usually complex with negative imaginary part for passive system, indicating a exponentially decaying wave in time with divergent amplitude in the far field, still there are special cases when $\text{Im}(\widetilde{\omega}_m) = 0$ like those in the bound states in the continuum (BIC) or the threshold of lasing with gain. The quality factor ($Q$-factor) of QNM can thus be defined as $Q_m = -\text{Re}(\widetilde{\omega}_m)/[2\text{Im}(\widetilde{\omega}_m)]$. $\mu_r(\widetilde{\omega}_m)$ and $\epsilon_r(\widetilde{\omega}_m)$ are the relative permeability and relative permittivity at the eigenfrequency, respectively. The material in this study is assumed to be non-magnetic, i.e., $\mu_r(\widetilde{\omega}_m) = 1$, while $\epsilon_r(\widetilde{\omega}_m)$ can be either dispersive or dispersionless. are the magnetic and electric field distribution of the $m^{\text{th}}$ QNM, and for convenience, the spatial quantity "$\mathbf{r}$" is often left out. Interestingly, for each QNM with eigenfrequency $\widetilde{\omega}_m$ and $[\tilde{\mathbf{H}}_m, \tilde{\mathbf{E}}_m]^{\text{T}}$, due to the Hermitian symmetry of $\epsilon_r(\widetilde{\omega}_m)$ ($\epsilon_r^*(\widetilde{\omega}_m) = \epsilon_r(-\widetilde{\omega}_m^*)$ and $\epsilon_r(\widetilde{\omega}_m^*) = \epsilon_r^*(-\widetilde{\omega}_m)$), "*" denotes conjugate operation, there exists another QNM with eigenfrequency $-\widetilde{\omega}_m^*$ and field distribution $[\tilde{\mathbf{H}}_m^*, \tilde{\mathbf{E}}_m^*]^{\text{T}}$, which can be proven easily by applying conjugate at both sides of Eq. (2) [8].

Besides, another group of solutions "inside" the open resonator satisfying Maxwell's equations with incoming wave (IWC) condition assuming sources far away:

$$\begin{bmatrix} 0 & -i(\mu_0\mu_r(-\widetilde{\omega}'_m))^{-1}\nabla \times \\ i(\epsilon_0\epsilon_r(-\widetilde{\omega}'_m))^{-1}\nabla \times & 0 \end{bmatrix} \begin{bmatrix} \tilde{\mathbf{H}}'_m(\mathbf{r}) \\ \tilde{\mathbf{E}}'_m(\mathbf{r}) \end{bmatrix} = -\widetilde{\omega}'_m \begin{bmatrix} \tilde{\mathbf{H}}'_m(\mathbf{r}) \\ \tilde{\mathbf{E}}'_m(\mathbf{r}) \end{bmatrix}, (3)$$

where the time harmonic term has the form of $\exp(i\widetilde{\omega}'_m t)$ considering the IWC

condition, and these equations are natural deduction of general time reverse symmetry [9]. These modes correspond to specific wave patterns injecting into the resonators without escape, i.e., they were totally absorbed, thus we hereafter denote them as the perfect absorbing modes (PAMs) [9, 10]. Especially, when $\text{Im}(\widetilde{\omega}'_m) = 0$, the open resonator can reach coherent perfect absorption (CPA) at a real frequency $\widetilde{\omega}'_m$ with a specific linear combination of inputs from different ports to synthesize $[\widetilde{\mathbf{H}}'_m, \widetilde{\mathbf{E}}'_m]^\text{T}$, and for single port open resonators, the CPA degenerates to common perfect absorption with independent inputs. From the Hermitian symmetry of $\epsilon_r(\widetilde{\omega}_m)$, it can be found that $\widetilde{\omega}'_m$ and $[\widetilde{\mathbf{H}}'_m, \widetilde{\mathbf{E}}'_m]$ of PAMs of a resonator are conjugate to those of the QNMs of its conjugate resonator with $\epsilon'_r(\mathbf{r},\omega) = \epsilon_r^*(\mathbf{r},\omega)$ (note that $\omega$ is not conjugated in the analytical dispersion formula), which can be understood in a picture of anti-lasing.

Obviously, for lossless passive resonators with time reversal symmetry, there are $\widetilde{\omega}'_m = \widetilde{\omega}_m^*$ and $[\widetilde{\mathbf{H}}'_m, \widetilde{\mathbf{E}}'_m]^\text{T} = [\widetilde{\mathbf{H}}_m^*, \widetilde{\mathbf{E}}_m^*]^\text{T}$, while for open resonators with lossy or gain material, it is difficult to find a quantitative relation between its QNMs and PAMs, but still it can be observed that they come in pairs and deviate gradually with the increasing of the degree of the time reversal symmetry breaking in the sense of perturbation. Yet when the *Q*-factor is high and QNMs are far away from each other, there remains $\text{Re}(\widetilde{\omega}_m) \approx \text{Re}(\widetilde{\omega}'_m)$ [2,9]. In fact, the QNMs are related with the poles of the scattering matrix (S matrix) and can be called perfect emitting modes (the resonator can be regarded as a transmitting antenna in this case), while the PAMs of the same system are related with the zeros of the S matrix (the resonator can be regarded as a receiving antenna in this case) [2,9], and they are equally important for the singularity representation of the S matrix. Furthermore, both the QNMs and PAMs can be regarded as special cases of more general scattering singularities (poles and zeros) of the scattering matrix [11, 12], and interesting anomalies can be inspected from the behavior of these singularities [13]. Specially, if the resonator is lossless and closed, the system described by Eqs. (2) is Hermitian, all $\widetilde{\omega}_m$ real, and QNMs and PAMs would degenerate to the orthogonal normal modes [14].

Furthermore, for periodic structures, the QNM also needs to satisfy the Bloch condition, and the fields have the form of $\tilde{\mathbf{H}}_m = \tilde{\mathbf{h}}_m \exp(i\mathbf{k}_b\mathbf{r})$ and $\tilde{\mathbf{E}}_m = \tilde{\mathbf{e}}_m \exp(i\mathbf{k}_b\mathbf{r})$ where $\tilde{\mathbf{h}}_m$ and $\tilde{\mathbf{e}}_m$ are the periodic parts of $\tilde{\mathbf{H}}_m$ and $\tilde{\mathbf{E}}_m$, respectively. The Bloch vector $\mathbf{k}_b$ is an independent argument, which is equal to the in-plane wave vector component of incident wave due to phase matching when coupled to the external excitation.

For resonators made of dispersionless material, Eqs. (2) is a standard linear eigen problem, which can be solved by mature algorithm [15] and even commercial software [16, 17]. But when the constitute material is dispersive, the eigen equations become nonlinear because the unknown $\tilde{\omega}_m$ enters the equations via $\epsilon_r(\tilde{\omega}_m)$. The easiest way to remove the nonlinearity is to fix $\epsilon_r(\tilde{\omega}_m)$ at a certain guess frequency (linearization point) around the interested eigenfrequency. This method is effective for weak dispersive material, but needs iteration to converge to the accurate eigenfrequency [2], and can only obtain one QNM at a time. Besides, the divergence property of QNM in the far field is also utilized to search poles one by one [18]. For problems which can obtain the S matrix in the complex frequency plane directly, like the Mie scattering and grating problem [19], the location of root method [3] or iterative method [20] can be used to find poles. Another group of method is to linearize the eigen equations by introducing appropriate auxiliary fields [3,8,21-24], which can solve all QNMs at a time in principle. Among them, the auxiliary fields (polarization **P** and current density **J**) introduced by Yan [3] and Gras [4] have the advantage of providing both clear physical meaning and semi- analytical mode excitation coefficients. For more details of methods to solve QNM, the reader are recommended with the seminar review by Lalanne et. al. [8,23] and Demesy et. al. [24]. Here we will adopt the auxiliary fields to linearize the eigen equations.

### 1.2. Partial-Fraction dispersion model and corresponding auxiliary fields

The optical dispersion of materials is related with the electronic band structure of constitute atoms/molecules, and can be described by appropriate models in different wavelength range [25]. For example, the dispersion of metal in the optical range can be modeled as Lorentz-Drude model [5] or Critical Point model [26] ; the dispersion of high- doping semiconductor in the THz range can be modeled as Drude model [6] ; the dispersion of organics and dielectrics in the ultraviolet range can be modeled as Gaussian model; the dispersion of dielectric in the visible range can be modeled by Sellmier equations [25] ;the dispersion of semiconductors around bandgap can be modeled as Tauc-Lorentz model [27] or Cody-Lorentz model [28] ; while the dispersion of polar liquid can be modeled as Debye model. The dispersion of material in the full range can be modeled as a superposition of different models, each contributing to specific range. These models have vivid physical meaning and satisfy the Hermitian symmetry, but not always fulfill the analyticity or obey the Kramers –Kronig (K-K) relation [25] . Moreover, it needs to select suitable models and elaborately fit parameters from experimental data for different materials in different range, which brings trouble in both constructing unified auxiliary fields or practical use in other cases, thus an unified dispersion model with as few fitting term as possible is highly required.

Garcia-Vergara et. al. [7] developed a unified Partial-fraction dispersion model to describe material dispersion, and proposed a relatively general algorithm to extract poles from experimental data. They start with the analyticity of permittivity assumption and express the permittivity as a rational function of ω, and extract the poles and zeros by a least square method. And finally the expression is converted to the following form according to the Mittag-Leffler Theorem [29, 30] and Hermitian symmetry:

$$\epsilon_r(\omega) = \epsilon_{r\infty}\left[1 + \sum_{j=1}^{N}\left(\frac{A_j}{\omega-\Omega_j} + \frac{-A_j^*}{\omega+\Omega_j^*}\right)\right], \quad (4)$$

where $\epsilon_{r\infty}$ is the real constant relative permittivity at high frequency, $(\Omega_j, -\Omega_j^*)$ are bigeminy poles of material oscillators, $(A_j, -A_j^*)$ are their complex amplitudes, and $N$ is the predefined truncation number. The Partial-Fraction dispersion model is analytical, obeys the K-K relation and Hermitian symmetry, and is not based on the phenomenological microscopic constitution of specific material, thus is a relatively universal dispersion model. Note that this model is essentially a variation of the so-called modified-Lorentz model [31, 32] which is able to fit the material dispersion with less terms and can thus reduce the computation task in algorithms like finite difference in time domain(FDTD) and also QNMEM in this study.

Meanwhile, the Partial-Fraction dispersion model is also compatible with several common used dispersion model like the Lorentz-Drude model ($\epsilon_r(\omega) = \epsilon_{r\infty} - \sum_{j=1}^{N} \frac{\omega_{pj}^2}{\omega^2 - \omega_0^2 + i\gamma_j \omega}$), the Critical Point model ($\epsilon_r(\omega) = \epsilon_{r\infty} + \sum_{j=1}^{N} A_j \Omega_j \left( \frac{e^{\phi_j}}{\Omega_j - \omega - i\Gamma_j} + \frac{e^{-\phi_j}}{\Omega_j + \omega + i\Gamma_j} \right)$), the Sellmier equations ($n^2 - 1 = A_s + \sum_{j=1}^{N} \frac{B_j \lambda^2}{\lambda^2 - \lambda_j^2}$), the Debye model ($\epsilon_r(\omega) = \epsilon_{r\infty} + \sum_{j=1}^{N} \frac{\Delta \epsilon_j}{1 - i\omega \tau_j}$), and the good conductors model ($\epsilon_r(\omega) = \epsilon_{r\infty} + \frac{i\sigma}{\omega \epsilon_0}$), the conversion relation of whom are listed in Tab. 1.

Table 1 Correspondence between several common models with the Partial-fraction dispersion model

| Parameters | $\epsilon_{r\infty}$ | Re ($\Omega_j$) | Im ($\Omega_j$) | Re ($A_j$) | Im ($A_j$) |
|---|---|---|---|---|---|
| Lorentz-Drude* | $\epsilon_{r\infty}$ | $\sqrt{\omega_{0j}^2 - (\gamma_j/2)^2}$ | $-\gamma_j/2$ | $\frac{-\omega_{pj}^2}{2\epsilon_{r\infty}\sqrt{\omega_{0j}^2 - (\gamma_j/2)^2}}$ | 0 |
| Critical Point | $\epsilon_{r\infty}$ | $\Omega_j$ | $-\Gamma_j$ | $-\frac{A_j \Omega_j \cos(\phi_j)}{\epsilon_{r\infty}}$ | $-\frac{A_j \Omega_j \sin(\phi_j)}{\epsilon_{r\infty}}$ |
| Sellmier | $1 + A_s$ | $2\pi c/\lambda_j$ | 0 | $\frac{-\pi c B_j}{(1+A_s)\lambda_j}$ | 0 |
| Debye | $\epsilon_{r\infty}$ | 0 | $-1/\tau_j$ | arbitary | $\frac{\Delta \epsilon_j}{2\epsilon_{r\infty}\tau_j}$ |
| Good conductor | $\epsilon_{r\infty}$ | 0 | 0 | arbitary | $\frac{\sigma}{2\epsilon_0 \epsilon_{r\infty}}$ |

*: Case for $\omega_{0j} > \gamma_j/2$

From Tab. 1, we can find that for Debye model there is $\Omega_j = -\Omega_j^*$ and $A_j = -A_j^*$, thus the pair of partial fractions is essentially degenerate; Beside, for Lorentz-Drude model with $\omega_{0j} < \gamma_j/2$ (it degenerates to Drude model when $\omega_{0j} = 0$), the two material poles become pure imaginary and do not fulfill Eq. (4), thus it needs special treatment. As each material pole fulfill the relation of Eq. (4) with itself, we can regard the two material poles as two pairs of degenerate poles, and derive the conversion relation in Tab. 2, where a and b denote two different pure imaginary material poles. Note that in real process each degenerate material poles is in fact merged to a single pole.

For the special case of Lorentz-Drude model with $\omega_{0j} = \gamma_j/2$), the material pole is no longer simple, and cannot be incorporated into the form of Eq. (3), but this situation is not common in practical so can be neglected.

Table 2 Conversion relation between the Lorentz-Drude model (case for $\omega_{0j} < \gamma_j/2$) and the Partial-Fraction dispersion model

| Parameters\Models | $\epsilon_{r\infty}$ | $\Omega_{ja}$ | $A_{ja}$ | $\Omega_{jb}$ | $A_{jb}$ |
|---|---|---|---|---|---|
| Lorentz-Drude | $\epsilon_{r\infty}$ | $i\left(-\frac{\gamma_j}{2} + \sqrt{\left(\frac{\gamma_j}{2}\right)^2 - \omega_{0j}^2}\right)$ | $\frac{i\omega_{pj}^2}{4\epsilon_{ro}\sqrt{(\gamma_j/2)^2 - \omega_{0j}^2}}$ | $i\left(-\frac{\gamma_j}{2} - \sqrt{\left(\frac{\gamma_j}{2}\right)^2 - \omega_{0j}^2}\right)$ | $-A_{ja}$ |

For nanoresonators with Partial-Fraction material dispersion, we will introduce auxiliary fields to linearize the eigen equations shown in Eq. (2). For each pair of material poles, we can define a pair of polarization vectors $\mathbf{P}_{1j}$ and $\mathbf{P}_{2j}$:

$$\mathbf{P}_{1j} = \frac{A_j \epsilon_0 \epsilon_{r\infty}}{\omega - \Omega_j}\mathbf{E}, \quad \mathbf{P}_{2j} = \frac{-A_j^* \epsilon_0 \epsilon_{r\infty}}{\omega + \Omega_j^*}\mathbf{E}, \tag{5}$$

where $\mathbf{P}_{1j}$ and $\mathbf{P}_{2j}$ also satisfy the following relation:

$$\omega(\mathbf{P}_{1j} + \mathbf{P}_{2j}) = (A_j - A_j^*)\epsilon_0 \epsilon_{r\infty}\mathbf{E} + \Omega_j \mathbf{P}_{1j} - \Omega_j^* \mathbf{P}_{2j} \tag{6}$$

In the following statement, for convenience, we only consider the case of one pair of material poles, but it is easy to extend the conclusion to multiple pairs of material poles. With the auxiliary fields defined in Eq. (5), we further define the augmented eigenvectors as

$$\widetilde{\mathbf{\Psi}}_m = \left[\widetilde{\mathbf{H}}_m, \widetilde{\mathbf{E}}_m, \widetilde{\mathbf{P}}_{1m}, \widetilde{\mathbf{P}}_{2m}\right]^{\mathrm{T}}, \tag{7}$$

thus Eq. (2) can be linearized as

$$\widehat{\mathbf{H}}\widetilde{\mathbf{\Psi}}_m = \begin{bmatrix} 0 & -i\mu_0^{-1}\nabla\times & 0 & 0 \\ i(\epsilon_0\epsilon_{r\infty})^{-1}\nabla\times & -(A_j - A_j^*) & -\Omega(\epsilon_0\epsilon_{r\infty})^{-1} & \Omega^*(\epsilon_0\epsilon_{r\infty})^{-1} \\ 0 & A\epsilon_0\epsilon_{r\infty} & \Omega & 0 \\ 0 & -A^*\epsilon_0\epsilon_{r\infty} & 0 & -\Omega^* \end{bmatrix}\widetilde{\mathbf{\Psi}}_m$$

$$= \widetilde{\omega}_m \widetilde{\mathbf{\Psi}}_m, \tag{8}$$

where $\widehat{\mathbf{H}}$ can be regarded as the Hamiltonian of the augmented Maxwell's equations. The auxiliary fields are only defined in dispersive domain, and is null in dispersionless domain where the Hamiltonian remains the same with that in Eq. (2). For the case of multiple material poles, it only needs to add corresponding rows and components.

### 1.3. Solving the QNMs

For some simple cases like Mie scattering of spheres or 1D Fabry-Pérot resonators,

we can obtain the analytical QNMs, but for most complicated resonators, Eqs. (2) and (8) usually do not have closed-form solutions, and we have to turn to numerical modeling to calculate the eigenvalue and eigenvectors. To implement the numerical modeling, we need to use some technique to truncate the original eigen equations defined in the infinite open space to map eigen equations in a finite closed domain, and at least the leading eigenvalues/eigenvectors in the unperturbed domain should remain unchanged and still satisfy the outgoing wave condition. One simple and elegant technique is to "wrap" the resonator with perfect matched layer (PML) to truncate the infinite domain while mimicking the OWC condition, and the PML with exterior surface of either perfect electric conductor (PEC) or perfect magnetic conductor (PMC) is thick enough to damp the field injecting into it [3,8] . For periodic structures, only the nonperiodic direction needs PML truncation while the boundary condition along periodic direction is still the Floquet-Bloch condition. PML nowadays is a very popular technique in computational electromagnetic [33, 34] , which is an impedance-matched virtual domain placed in the outer of the physical domain as a "light trap" and does not cause any reflection in principle. PML can be implemented by field decomposition method, complex coordinate stretch method or anisotropic material method, which are equivalent. Although the advanced complex coordinate stretch method is most used in current commercial softwares, the anisotropic material method is much easier to understand and implement manually. In Cartesian coordinate system, PML can be regarded as a domain made of anisotropic material with following permittivity tensor $\bar{\bar{\epsilon}}$ and permeability tensor $\bar{\bar{\mu}}$:

$$\bar{\bar{\epsilon}} = \epsilon \begin{bmatrix} \frac{s_y s_z}{s_x} & 0 & 0 \\ 0 & \frac{s_x s_z}{s_y} & 0 \\ 0 & 0 & \frac{s_x s_y}{s_z} \end{bmatrix}, \quad \bar{\bar{\mu}} = \mu \begin{bmatrix} \frac{s_y s_z}{s_x} & 0 & 0 \\ 0 & \frac{s_x s_z}{s_y} & 0 \\ 0 & 0 & \frac{s_x s_y}{s_z} \end{bmatrix}, \quad (9)$$

where $\epsilon$ and $\mu$ are the permittivity and permeability of domain truncated by the PML. $s_x$, $s_y$ and $s_z$ are the stretch factors along $x$, $y$ and $z$, respectively, which are complex constants or functions of spatial locations [34]. For this study, we can simply choose these stretch factors as complex constants with positive real parts, and to fulfill the

OWC, the imaginary part of stretch factors should be positive which makes the PML absorbing. On the other hand, to solve the PAMs, the problem can be transformed to solve the QNMs of the conjugate problem, or equivalently, we can use the conjugate PML (cPML) with negative imaginary part of stretch factors for Eq. (8) to mimic IWC condition. Another advantage of using PML is that in PML the QNM decays exponentially rather than diverge, making it square-integrable in the mapped space and also making the normalization possible [8,36]. The function of the PML is depicted in Fig. 1. The discussion hereafter all aims at the eigen problem truncated by PML.

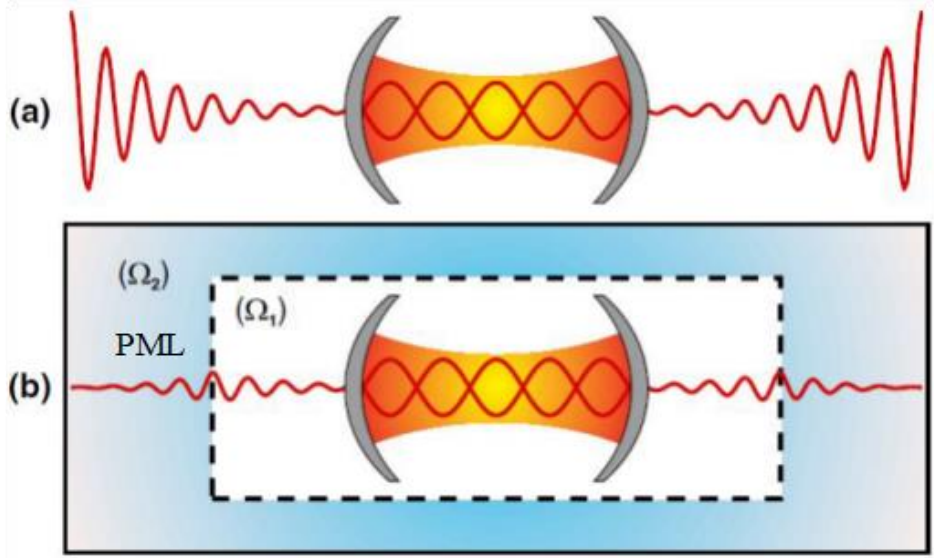

Fig. 1 (a) QNM of an open resonator. The field of QNM keeps finite "in" the cavity and decays in the near field "outside" of the cavity, but increases exponentially at a certain distance ($\sim Q\lambda_m$). (b) QNM in the open resonator truncated by PML. The field of QNM nearly does not change in the physic domain (Domain $\Omega_1$), but decays exponentially in PML (Domain $\Omega_2$), and satisfies the OWC [36] .

After the PML mapping and discretization, the original continuous eigen equations are converted to a discretized operator defined in finite closed domain including PML. However, this transformation is only valid in a finite complex frequency range $\mathcal{F}$, and thus only the eigenvalues/eigenvectors of mapped $\widehat{\mathbf{H}}$ in $\widetilde{\omega}_m \in \mathcal{F}$ correspond to original QNMs, and all the other modes are collectively called the PML modes. The PML modes can roughly be classified into two classes [1,3,8,37]. The first class corresponds the QNMs of the original system which are not accurately resolved by the mapped system, while the second class results from the continuum of the original problem which are rotated from the real frequency axis to the complex plane [1,3,37] . For scatters in homogenous background, the eigenfrequencies of the second class QNMs are distributed in an inclined line trough the origin. But for scatters on slabs or substrates [3] , or for gratings [1,37] , the eigenfrequencies of the second class QNMs are distributed in several branches due to the emergence of guide mode resonance [3] or higher diffraction orders [1,37] , then the QNMs and the PML modes can be "entangled" and difficult to distinguish easily.

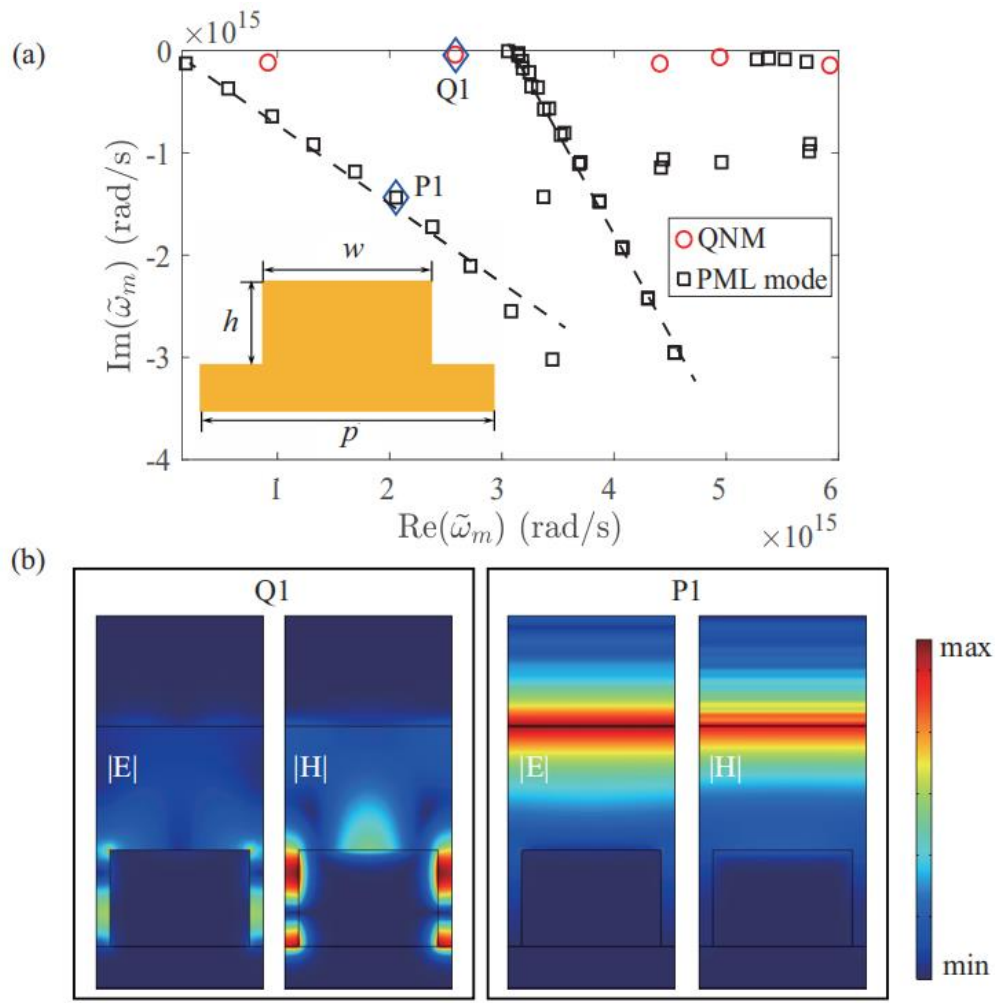

Fig. 2 (a) QNMs and PML modes of a periodic structure. The two rhombus denote a QNM Q1 and a PML mode P1, respectively. The PML modes around the two dash lines corresponds to the discretized diffraction order continuum. The inset shows a unit cell of the grating. (b) Normalized electric field |E| and magnetic field |H| of Q1 and P1. The PML has a thickness of 400 nm with stretch factors $s_x = 1$ and $s_z = 3 + 3i$.

Figure 2(a) shows part of the eigenfrequencies of a periodic structure, where two apparent straight branches can be figured out. Generally speaking, the PML modes varies with the change of PML parameters, but the QNM would not, and we can utilize this property to distinguish them. With this technique, 5 QNMs are recognized in Fig. 2(a). Note that numerically this identification operation is not necessary because both QNMs and PML modes are important for the full field reconstruction due to the completeness assumption [3,8].

The eigen equations shown in Eq. (8) can be solved efficiently by mature linear algorithm like the first companion linearization [15], which is also implemented in commercial softwares like COMSOL Multiphysics [16]. In this study, we will develop our own solver in the basis of eigen solver in COMSOL Multiphysics to save effort. To accommodate the setting of the built-in eigen solver, we need to transform Eq. (8) into the form below:

$$\widehat{\mathbf{K}}\begin{bmatrix}\widetilde{\mathbf{E}}_m\\\widetilde{\mathbf{P}}_{1m}\\\widetilde{\mathbf{P}}_{2m}\end{bmatrix} + \widetilde{\omega}_m\widehat{\mathbf{C}}\begin{bmatrix}\widetilde{\mathbf{E}}_m\\\widetilde{\mathbf{P}}_{1m}\\\widetilde{\mathbf{P}}_{2m}\end{bmatrix} + \widetilde{\omega}_m^2\widehat{\mathbf{M}}\begin{bmatrix}\widetilde{\mathbf{E}}_m\\\widetilde{\mathbf{P}}_{1m}\\\widetilde{\mathbf{P}}_{2m}\end{bmatrix} = 0 \quad (10)$$

where $\widehat{\mathbf{K}}$, $\widehat{\mathbf{C}}$ and $\widehat{\mathbf{C}}$ are the stiffness matrix, the damping matrix and the mass matrix, respectively, and are defined as follows:

$$\widehat{\mathbf{K}} = \begin{bmatrix}\nabla\times\mu_0^{-1}\nabla\times & 0 & 0\\-A\epsilon_0\epsilon_{r\infty} & -\Omega & 0\\A^*\epsilon_0\epsilon_{r\infty} & 0 & \Omega^*\end{bmatrix}, \widehat{\mathbf{C}} = \begin{bmatrix}-(A-A^*)\epsilon_0\epsilon_{r\infty} & -\Omega & \Omega^*\\0 & 1 & 0\\0 & 0 & 1\end{bmatrix},$$

$$\widehat{\mathbf{M}} = \begin{bmatrix}-\epsilon_0\epsilon_{r\infty} & 0 & 0\\0 & 0 & 0\\0 & 0 & 0\end{bmatrix} \quad (11)$$

In the COMSOL Multiphysics environment, Eq. (9) should be transformed into weak forms and imported into the built-in solver to output the ultimate eigenfrequencies/eigenvectors of QNMs or PML modes [3].

To demonstrate the performance of the QNM solver, here we calculate the eigenfrequencies/eigenvectors of a 1D periodic structure. The 1D periodic structure is a indention metal grating on metal substrate in air, as is shown in the inset of Fig. 2(a). The dispersion of the metal can be modeled by a pair of partial fraction with $\epsilon_{r\infty} = 1.5$, $A = (-3\times 10^{17} + i2\times 10^{14})\ rad/s, \Omega = (1.5\times 10^{14} - i5\times 10^{13})\ rad/s$ . The period $p$=600 nm, the width is $w$=500 nm, and the height of the grating is $h$=350 nm. In a personal computer (Win10 64bit system with Intel i5 CPU, 16 GB memory, and dominant frequency of 3.2 GHz), the time needed to solve a single eigen mode is about 2 seconds, which is on the same level of full wave simulation at single frequency point. Figure 1.2(a) only shows part of the eigenfrequencies with Re(m ) > 0 and for $\mathbf{k_b} = 0$, and two inclined dash lines depict the second class of PML modes [1]. Figure 2(a) also indicates that QNMs are located close to the real axis with higher Q-factor. Figure 2(b) shows the normalized electric and magnetic field distribution of QNM Q1 and PML mode P1, which clearly indicates that the field of QNM mainly concentrate around the physical structure, while the field of PML modes concentrate around the PML domain.

## 2. Quasi-normal mode expansion method

Having obtained the eigenmodes (eigenfrequencies and eigenvectors) , we can decompose the electromagnetic field into the linear combination of eigenmodes. Although QNMs and PML modes are equally important numerically, we still call this method Quasi-normal mode expansion method (QNMEM) considering the dominant contribution of QNMs. The key to the QNMEM is the weight of each eigenmode, or the excitation coefficients of each eigenmode, which can be acquired via orthogonal decomposition method or the residue method [8] . In this study, we adopt the orthogonal decomposition method to derive the closed-form expression of excitation coefficients utilizing the bi-orthogonality of eigenmodes and normalization of eigenmodes.

### 2.1. Bi-orthogonality, normalization and completeness of eigenmodes

The bi-orthogonality and normalization are based on the Unconjugated form of the Lorentz reciprocity theorem (see Supplement 1). For the scattering of aperiodic structure, considering two eigenmodes of the PML mapped resonators, $\{\widetilde{\omega}_m, \widetilde{\Psi}_m\}$ and $\{\widetilde{\omega}_n, \widetilde{\Psi}_n\}$, they both satisfy the source-free Maxwell's equations $\hat{\mathbf{H}}\widetilde{\Psi}_m = \widetilde{\omega}_m\widetilde{\Psi}_m$ and $\hat{\mathbf{H}}\widetilde{\Psi}_n = \widetilde{\omega}_n\widetilde{\Psi}_n$. Taking them into Eq. (S1-5) leads to

$$(\widetilde{\omega}_m - \widetilde{\omega}_n) \iiint_V \widetilde{\boldsymbol{\Psi}}_m^{\mathrm{T}} \cdot \widehat{\mathbf{D}} \widetilde{\boldsymbol{\Psi}}_n d^3\mathbf{r} = i \iint_\Sigma \left( \widetilde{\mathbf{E}}_m \times \widetilde{\mathbf{H}}_n - \widetilde{\mathbf{E}}_n \times \widetilde{\mathbf{H}}_m \right) \cdot d\mathbf{s}, \qquad (12)$$

where the integration domain $V$ is the full PML mapped space (including the PML), and the exterior $\Sigma$ of PML is set as PEC or PMC which can the tangential components of electric or magnetic field null, making the right-hand side of Eq. (11) zero. Thus when $m \neq n$ and $\widetilde{\omega}_m \neq \widetilde{\omega}_n$, there is $\iiint_V \widetilde{\boldsymbol{\Psi}}_m^{\mathrm{T}} \cdot \widehat{\mathbf{D}} \widetilde{\boldsymbol{\Psi}}_n d^3\mathbf{r} = 0$, while when $m = n$, the normalization of the eigenmode can be implemented by scaling the mode field to fulfill $\iiint_V \widetilde{\boldsymbol{\Psi}}_m^{\mathrm{T}} \cdot \widehat{\mathbf{D}} \widetilde{\boldsymbol{\Psi}}_m d^3\mathbf{r} = 1$. Eventually we can get the relation for all eigenmodes (both QNMs and PML modes) below

$$\iiint_V \widetilde{\boldsymbol{\Psi}}_m^{\mathrm{T}} \cdot \widehat{\mathbf{D}} \widetilde{\boldsymbol{\Psi}}_n d^3\mathbf{r} = \iiint_V \left( \widehat{\mathbf{D}}^* \widetilde{\boldsymbol{\Psi}}_m^* \right)^\dagger \cdot \widetilde{\boldsymbol{\Psi}}_n d^3\mathbf{r} = \delta_{mn}, \qquad (13)$$

where "†" denotes conjugate transpose, and $\delta_{mn}$ is Kronecker delta. Equation (13) indicates that $\widehat{\mathbf{D}}^* \widetilde{\boldsymbol{\Psi}}_m^*$ is the adjoint eigenmode of $\widetilde{\boldsymbol{\Psi}}_n$ and they constitute a group bi-orthogonal basis.

In previous research[8,36,38], the form of the normalization of PML-regularized QNMs is $\iiint_V \left[ \left.\frac{\partial \omega \epsilon}{\partial \omega}\right|_{\widetilde{\omega}_m} \widetilde{\mathbf{E}}_m \cdot \widetilde{\mathbf{E}}_m - \left.\frac{\partial \omega \mu}{\partial \omega}\right|_{\widetilde{\omega}_m} \widetilde{\mathbf{H}}_m \cdot \widetilde{\mathbf{H}}_m \right] = 1$, which proves to be more stable and efficient than several other normalization method as is detailed in Ref. [35]. It can be verified that it is equivalent to Eq. (13).

We need to put some emphasis on the possible degenerate states, i.e. two eigenmodes with the same eigenfrequencies ($m \neq n$ but $\widetilde{\omega}_m = \widetilde{\omega}_n$). For some special case, for example, the polarization degenerate modes of structures with certain symmetry, it is easy to prove that Eq. (13) still applies due to the orthogonality of eigenmodes. But for some other complex cases, like the so-called exponential point, both the eigenfrequencies and eigenvectors are degenerate, the applicability of Eq. (13) needs further investigation.

For the diffraction of periodic structure, the form of the bi-orthogonality and normalization of eigenmodes is a little different, because only the nonperiodic direction is wrapped by PML with PEC/PMC, while the boundary condition along the periodic directions are Floquet-Bloch condition. Therefore, an auxiliary eigenmode is introduced. For each eigenmode $\{\widetilde{\omega}_{\mathbf{k}_b,m}, \widetilde{\boldsymbol{\Psi}}_{\mathbf{k}_b,m}\}$ satisfying $\widehat{\mathbf{H}}_{\mathbf{k}_b} \widetilde{\boldsymbol{\Psi}}_{\mathbf{k}_b,m} =$

$\widetilde{\omega}_{\mathbf{k_b},m}\widetilde{\mathbf{\Psi}}_{\mathbf{k_b},m}$, there exists another eigenmode [23,39] $\{\widetilde{\omega}_{-\mathbf{k_b},m}, \widetilde{\mathbf{\Psi}}_{-\mathbf{k_b},m}\}$ satisfying $\widehat{\mathbf{H}}_{-\mathbf{k_b}}\widetilde{\mathbf{\Psi}}_{-\mathbf{k_b},m} = \widetilde{\omega}_{-\mathbf{k_b},m}\widetilde{\mathbf{\Psi}}_{-\mathbf{k_b},m}$ and $\widetilde{\omega}_{\mathbf{k_b},m} = \widetilde{\omega}_{-\mathbf{k_b},m}$. The Bloch vector for these two eigenmodes are $\mathbf{k_b}$ and $-\mathbf{k_b}$, respectively. There are

$$\widetilde{\mathbf{\Psi}}_{\mathbf{k_b},m} = \left[\tilde{\mathbf{h}}_{\mathbf{k_b},m}, \tilde{\mathbf{e}}_{\mathbf{k_b},m}, \tilde{\mathbf{p}}_{1,\mathbf{k_b},m}, \tilde{\mathbf{p}}_{2,\mathbf{k_b},m}\right]^{\mathrm{T}} \exp(i\mathbf{k_b}\mathbf{r}), \tag{14a}$$

$$\widetilde{\mathbf{\Psi}}_{-\mathbf{k_b},m} = \left[\tilde{\mathbf{h}}_{-\mathbf{k_b},m}, \tilde{\mathbf{e}}_{-\mathbf{k_b},m}, \tilde{\mathbf{p}}_{1,-\mathbf{k_b},m}, \tilde{\mathbf{p}}_{2-\mathbf{k_b},m}\right]^{\mathrm{T}} \exp(-i\mathbf{k_b}\mathbf{r}), \tag{14b}$$

where $\tilde{\mathbf{h}}_{\pm\mathbf{k_b},m}, \tilde{\mathbf{e}}_{\pm\mathbf{k_b},m}, \tilde{\mathbf{p}}_{1,\pm\mathbf{k_b},m}$ and $\tilde{\mathbf{p}}_{2,\pm\mathbf{k_b},m}$ are all periodic functions. Replacing the $\widetilde{\mathbf{\Psi}}_m$ and e $\widetilde{\mathbf{\Psi}}_n$ in Eq. (11) with $\widetilde{\mathbf{\Psi}}_{\mathbf{k_b},m}$ and $\widetilde{\mathbf{\Psi}}_{-\mathbf{k_b},n}$, $\exp(i\mathbf{k_b}\mathbf{r})$ and $\exp(-i\mathbf{k_b}\mathbf{r})$ cancel each other, making integral in the right-hand side of Eq. (12) a periodic function and the integration on the exterior boundary zero. Thus the bi-orthogonality and normalization relation for the eigenmodes of periodic structure is

$$\iiint_V \widetilde{\mathbf{\Psi}}_{-\mathbf{k_b},m}^{\mathrm{T}} \cdot \widehat{\mathbf{D}}\widetilde{\mathbf{\Psi}}_{\mathbf{k_b},n} d^3\mathbf{r} = \iiint_V \left(\widehat{\mathbf{D}}^*\widetilde{\mathbf{\Psi}}_{-\mathbf{k_b},m}^*\right)^{\dagger} \cdot \widetilde{\mathbf{\Psi}}_{\mathbf{k_b},n} d^3\mathbf{r} = \delta_{mn}, \tag{15}$$

which also indicates that $\widehat{\mathbf{D}}^*\widetilde{\mathbf{\Psi}}_{-\mathbf{k_b},m}^*$ is the adjoint eigenmode of $\widetilde{\mathbf{\Psi}}_{\mathbf{k_b},m}$ and they also constitute a group bi-orthogonal basis. Equations (13) and (15) also apply to multiple pairs of partial fractions dispersion cases.

Except for the special case of $\mathbf{k_b} = 0$, $\widetilde{\mathbf{\Psi}}_{-\mathbf{k_b},m}$ usually needs a re-computation. But for centrosymmetric structures (C$_{2v}$), i.e., $\epsilon(-x,-y) = \epsilon(x,y)$ (suppose the periodic directions are in the $xy$ plane), $\widetilde{\mathbf{\Psi}}_{-\mathbf{k_b},m}$ can be inferred from $\widetilde{\mathbf{\Psi}}_{\mathbf{k_b},m}$, i.e., $\widetilde{\mathbf{\Psi}}_{-\mathbf{k_b},m,x}(x,y,z) = -\widetilde{\mathbf{\Psi}}_{\mathbf{k_b},m,x}(-x,-y,z), \widetilde{\mathbf{\Psi}}_{-\mathbf{k_b},m,y}(x,y,z) = -\widetilde{\mathbf{\Psi}}_{\mathbf{k_b},m,y}(-x,-y,z)$ and $\widetilde{\mathbf{\Psi}}_{-\mathbf{k_b},m,z}(x,y,z) = \widetilde{\mathbf{\Psi}}_{\mathbf{k_b},m,z}(-x,-y,z)$. Note that even though all the conclusions are derived for 2D periodic structures, it also applies to 1D periodic structures.

To guarantee the strictness of the spectral decomposition, the completeness of the eigenmodes is required. To analyze the completeness of the eigenmodes of the Maxwell operator $\widehat{\mathbf{H}}$, we first need to introduce its adjoint eigenmodes [1,40], i.e., the eigenmodes of $\widehat{\mathbf{H}}^{\dagger}$. For aperiodic structures, applying conjugate to both sides of Eq. (8) gives $\widehat{\mathbf{H}}^*\widetilde{\mathbf{\Psi}}_m^* = \widetilde{\omega}_m^*\widetilde{\mathbf{\Psi}}_m^*$, together with $\widehat{\mathbf{H}}^{\dagger}\widehat{\mathbf{D}}^* = \widehat{\mathbf{D}}^*\widehat{\mathbf{H}}^*$ it outputs

$$\widehat{\mathbf{H}}^{\dagger}\widehat{\mathbf{D}}^{*}\widetilde{\mathbf{\Psi}}_{m}^{*} = \widetilde{\omega}_{m}^{*}\widehat{\mathbf{D}}^{*}\widetilde{\mathbf{\Psi}}_{m}^{*}, \tag{16}$$

which indicates that $\widehat{\mathbf{D}}^{*}\widetilde{\mathbf{\Psi}}_{m}^{*}$ is the adjoint eigenmode of $\widetilde{\mathbf{\Psi}}_{m}$, and they constitute a group of bi-orthogonal basis. However, as for the completeness of this group of bi-orthogonal basis in mathematical, except for the simple cases like F-P cavity[8] and Mie scattering problem [41] which have been tested easily inside the resonators, it is tough to verify all general cases one by one. Thus we have to assume it is complete from the perspective that the reconstructed results are usually consistent with experiment, and get the relation below

$$\sum_{m=1}^{\infty} \widetilde{\mathbf{\Psi}}_{m}(\mathbf{r}') \cdot \left[\widehat{\mathbf{D}}^{*}\widetilde{\mathbf{\Psi}}_{m}^{*}(\mathbf{r})\right]^{\dagger} = \sum_{m=1}^{\infty} \widetilde{\mathbf{\Psi}}_{m}(\mathbf{r}') \cdot \widetilde{\mathbf{\Psi}}_{m}^{T}(\mathbf{r})\widehat{\mathbf{D}} = \widehat{\mathbf{I}}\delta(\mathbf{r}-\mathbf{r}'), \tag{17}$$

where $\widehat{\mathbf{I}}$ is unit matrix which has the same dimension with $\mathbf{D}$.

Remember that we have already inferred from Eq. (15) that $\widehat{\mathbf{D}}^{*}\widetilde{\mathbf{\Psi}}_{-\mathbf{k}_{b},m}^{*}$ is the adjoint eigenmode of $\widetilde{\mathbf{\Psi}}_{\mathbf{k}_{b},m}$ and they also span a group bi-orthogonal basis. And when assuming the completeness alike, we can also obtain that

$$\sum_{m=1}^{\infty} \widetilde{\mathbf{\Psi}}_{\mathbf{k}_{b},m} \cdot (\mathbf{r}')\left[\widehat{\mathbf{D}}^{*}\widetilde{\mathbf{\Psi}}_{-\mathbf{k}_{b},m}^{*}(\mathbf{r})\right]^{\dagger} = \sum_{m=1}^{\infty} \widetilde{\mathbf{\Psi}}_{\mathbf{k}_{b},m}(\mathbf{r}') \cdot \widetilde{\mathbf{\Psi}}_{-\mathbf{k}_{b},m}^{T}(\mathbf{r})\widehat{\mathbf{D}}$$
$$= \widehat{\mathbf{I}}\delta(\mathbf{r}-\mathbf{r}') \tag{18}$$

### 2.2. Excitation coefficients

Due to the completeness and bi-orthogonality of the eigenmodes, we can expand the scattered/diffracted field into the linear superposition of normalized QNMs and PML modes. The weightiness of each eigenmode, or the excitation coefficient of each eigenmode, characterizes the contribution of each eigenmode to full field, and should be related with the similarity of eigenmodes and incident filed in both space (eigenmode field distribution) and time/frequency (eigenfrequency) intuitively.

For the aperiodic structure in Fig. 3(a) and periodic structure in 1.3(b), the permittivity of the whole structure can be denoted as $\epsilon = \epsilon_{0}(\epsilon_{\mathrm{rbg}} + \Delta\epsilon_{r})$, where $\epsilon_{\mathrm{rbg}}$ is the background permittivity (not necessarily homogeneous) and $\Delta\epsilon_{r}$ is null outside the resonator domain $V_{\mathrm{res}}$. Incident field with real frequency $\omega$ can be generated by external current source $J_E$ or magnetic source $J_M$ located either in the near field or in the far field [42], in this study, we are only concerned on external source in the far field. For periodic structures, the incident filed is usually plane wave which can be regarded as wave generated by $J_E$ or $J_M$ at infinity. The total field with auxiliary field is defined as $\Psi_{total} = [H_{total}, E_{total}, P_{1total}, P_{2total}]^T$, and satisfies the following augmented Maxwell's equations[3,42]

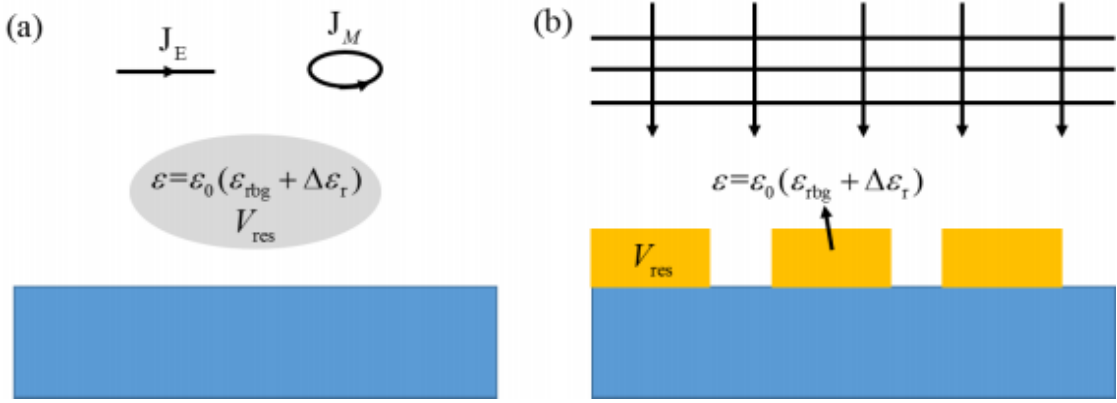

Fig. 3 (a) Scattering problem of aperiodic structure. $J_E$ and $J_M$ are current source and magnetic source, respectively. (b) Diffraction problem of periodic structure with plane wave incidence.

$$\hat{H}\Psi_{total} = \omega\Psi_{total} + \begin{bmatrix} i\mu_0^{-1}J_M \\ i(\epsilon_0\epsilon_{r\infty})^{-1}J_E \\ 0 \\ 0 \end{bmatrix}, \quad (19)$$

where $\hat{H}$ has the same form with the one in Eq. (8).

The background field is defined as $\Psi_{bg} = [H_{bg}, E_{bg}, P_{1bg}, P_{2bg}]^T$, and is null in the resonator domain $V_{res}$, i.e., when $\mathbf{r} \in V_{res}$, $\Psi_{bg} = [H_{bg}, E_{bg}, 0, 0]^T$. The background field satisfies

$$\hat{H}\Psi_{bg} = \omega\Psi_{bg} + \begin{bmatrix} i\mu_0^{-1}J_M \\ i(\epsilon_0\epsilon_{r\infty})^{-1}J_E \\ 0 \\ 0 \end{bmatrix} - S_{bg}, \quad (20)$$

where $S_{bg} = 0$ where $\mathbf{r} \notin V_{res}$. For $\mathbf{r} \in V_{res}$, there is (see Supplement 3 for details)

$$S_{bg} = [0, [\omega(\epsilon_{r\infty} - \epsilon_{rbg})/\epsilon_{r\infty} + (A - A^*)]E_{bg}, -A\epsilon_0\epsilon_{r\infty}E_{bg}, A^*\epsilon_0\epsilon_{r\infty}E_{bg}]^T. (21)$$

For the material dispersion case of multiple partial fractions, there is

$$\mathbf{S}_{\text{bg}} = \left[0, \left[\omega(\epsilon_{r\infty} - \epsilon_{\text{rbg}})/\epsilon_{r\infty} + \sum_{j=1}^{N}(A_j - A_j^*)\right]\mathbf{E}_{\text{bg}}, -A_1\epsilon_0\epsilon_{r\infty}\mathbf{E}_{\text{bg}}, A_1^*\epsilon_0\epsilon_{r\infty}\mathbf{E}_{\text{bg}},\right.$$
$$\left.\cdots, -A_j\epsilon_0\epsilon_{r\infty}\mathbf{E}_{\text{bg}}, A_j^*\epsilon_0\epsilon_{r\infty}\mathbf{E}_{\text{bg}}\right]^{\text{T}}. \tag{22}$$

Subtracting Eq. (19) by Eq. (20), and from the definition of scattered field $\boldsymbol{\Psi}_{\text{sca}} = \boldsymbol{\Psi}_{\text{total}} - \boldsymbol{\Psi}_{\text{bg}}$, there is

$$\widehat{\mathbf{H}}\boldsymbol{\Psi}_{\text{sca}} = \omega\boldsymbol{\Psi}_{\text{sca}} + \mathbf{S}_{\text{bg}}, \tag{24}$$

which indicates that $\mathbf{S}_{\text{bg}}$ is the source of the scattered field.

The diffraction of periodic structure can be regarded as a special scattering problem, it only needs to replace the $\boldsymbol{\Psi}_{\text{sca}}$ in Eq. (21) with $\boldsymbol{\Psi}_{\mathbf{k}_b,\text{diff}}$ (the total field deducting the background filed where the background field is not necessary to be the incident field), while the form of $\mathbf{S}_{\text{bg}}$ remains unchanged.

The scattered field of aperiodic structure $\boldsymbol{\Psi}_{\text{sca}}(\mathbf{r},\omega)$ can expanded in the whole PML mapped space as

$$\boldsymbol{\Psi}_{\text{sca}}(\mathbf{r},\omega) = \sum_{m=1}^{\infty} \Lambda_m(\omega)\widetilde{\boldsymbol{\Psi}}_m(\mathbf{r}), \tag{25}$$

where $\Lambda_m(\omega)$ is the excitation coefficients of a certain eigenmode. Applying $\iiint_V d^3\mathbf{r}\widetilde{\boldsymbol{\Psi}}_m^{\text{T}}\widehat{\mathbf{D}}$ to both sides of Eq. (24), substituting Eq. (25) into it, and considering the bi-orthogonality defined in Eq. (13) gives

$$\begin{aligned}\Lambda_m(\omega) &= \frac{\iiint_{V_{\text{res}}} \widetilde{\boldsymbol{\Psi}}_m^{\text{T}} \cdot \widehat{\mathbf{D}}_{\text{bg}} d^3\mathbf{r}}{\widetilde{\omega}_m - \omega} \\ &= \frac{\widetilde{\omega}_m}{\widetilde{\omega}_m - \omega}\left\langle \widetilde{\mathbf{E}}_m^*(\mathbf{r}) \middle| \epsilon_0[\epsilon_r(\widetilde{\omega}_m,\mathbf{r}) - \epsilon_{\text{rbg}}(\omega,\mathbf{r})] \middle| \mathbf{E}_{\text{bg}}(\omega,\mathbf{r})\right\rangle_{V_{\text{res}}} \\ &\quad + \left\langle \widetilde{\mathbf{E}}_m^*(\mathbf{r}) \middle| \epsilon_0[\epsilon_{\text{rbg}}(\omega,\mathbf{r}) - \epsilon_{r\infty}(\mathbf{r})] \middle| \mathbf{E}_{\text{bg}}(\omega,\mathbf{r})\right\rangle_{V_{\text{res}}} \\ &= \frac{\widetilde{\omega}_m}{\widetilde{\omega}_m - \omega}\iiint_{V_{\text{res}}} \epsilon_0[\epsilon_r(\widetilde{\omega}_m,\mathbf{r}) - \epsilon_{\text{rbg}}(\omega,\mathbf{r})]\widetilde{\mathbf{E}}_m(\mathbf{r}) \cdot \mathbf{E}_{\text{bg}}(\omega,\mathbf{r})d^3\mathbf{r} \\ &\quad + \iiint_{V_{\text{res}}} \epsilon_0[\epsilon_{\text{rbg}}(\omega,\mathbf{r}) - \epsilon_{r\infty}(\mathbf{r})]\widetilde{\mathbf{E}}_m(\mathbf{r}) \cdot \mathbf{E}_{\text{bg}}(\omega,\mathbf{r})d^3\mathbf{r},\end{aligned} \tag{26}$$

where $\left\langle \widetilde{\mathbf{E}}_m^*(\mathbf{r}) \middle| f(\mathbf{r}) \middle| \mathbf{E}_{\text{bg}}(\omega,\mathbf{r})\right\rangle_{V_{\text{res}}} = \iiint_{V_{\text{res}}} f(\mathbf{r})\widetilde{\mathbf{E}}_m(\mathbf{r}) \cdot \mathbf{E}_{\text{bg}}(\omega,\mathbf{r})d^3\mathbf{r}$ is the classical notation of overlap integration [11] with $f(\mathbf{r})$ the weighing function.

Equation (26) also applies to the case of multiple partial fractions. The closed-form expression of excitation coefficient provides great convenience for field reconstruction, phenomenon analysis and inverse design of resonant nanostructures. For the diffracted field $\Psi_{k_b,\text{diff}}$ of the periodic resonators, the expression of expansion and excitation coefficients are

$$\Psi_{k_b,\text{dif}}(\mathbf{r},\omega) = \sum_{m=1}^{\infty} \Lambda_{k_b,m}(\omega)\widetilde{\Psi}_{k_b,m}(\mathbf{r}) \tag{27}$$

$$\begin{aligned}
\Lambda_{k_b,m}(\omega) &= \frac{\iiint_{V_{\text{res}}} \widetilde{\Psi}_{-k_b,m}^{\text{T}} \cdot \hat{\mathbf{D}} \mathbf{S}_{\text{bg}} d^3\mathbf{r}}{\widetilde{\omega}_m - \omega} \\
&= \frac{\widetilde{\omega}_m}{\widetilde{\omega}_m - \omega} \left\langle \widetilde{\mathbf{E}}_{-k_b,m}^*(\mathbf{r}) \middle| \epsilon_0[\epsilon_r(\widetilde{\omega}_m,\mathbf{r}) - \epsilon_{\text{rgg}}(\omega,\mathbf{r})] \middle| \mathbf{E}_{\text{bg}}(\omega,\mathbf{r}) \right\rangle_{V_{\text{res}}} \\
&\quad + \left\langle \widetilde{\mathbf{E}}_{-k_b,m}^*(\mathbf{r}) \middle| \epsilon_0[\epsilon_{\text{rbg}}(\omega,\mathbf{r}) - \epsilon_{\text{ro}}(\mathbf{r})] \middle| \mathbf{E}_{\text{bg}}(\omega,\mathbf{r}) \right\rangle_{V_{\text{res}}} \\
&= \frac{\widetilde{\omega}_m}{\widetilde{\omega}_m - \omega} \iiint_{V_{\text{res}}} \epsilon_0[\epsilon_r(\widetilde{\omega}_m,\mathbf{r}) - \epsilon_{\text{rbg}}(\omega,\mathbf{r})] \widetilde{\mathbf{E}}_{-k_b,m}(\mathbf{r}) \cdot \mathbf{E}_{\text{bg}}(\omega,\mathbf{r}) d^3\mathbf{r} \\
&\quad + \iiint_{V_{\text{res}}} \epsilon_0[\epsilon_{\text{rbg}}(\omega,\mathbf{r}) - \epsilon_{\text{roo}}(\mathbf{r})] \widetilde{\mathbf{E}}_{-k_b,m}(\mathbf{r}) \cdot \mathbf{E}_{\text{bg}}(\omega,\mathbf{r}) d^3\mathbf{r}
\end{aligned} \tag{28}$$

**2.3. Absorption/scattering/extinction cross section**

For the scattering problem of aperiodic structure, we can derive the absorption/scattering/extinction cross section according the Poynting theorem (see Supplement 2). At a certain real frequency $\omega$, Eq. (S2-3) becomes

$$P_{\text{ext}} = P_{\text{abs}} + P_{\text{sca}}, \tag{29}$$

where $P_{\text{ext}}$ is the extinction power which is actually the incident power $P_{\text{inp}}$ in Eq. (S2-3); $P_{\text{abs}}$ is the absorption power; $P_{\text{sca}}$ is the scattering power which is actually the radiation power $P_{\text{rad}}$ in Eq. (S2-3). These quantities can be obtained from Eqs. (S2-4b) ~ (S2-4d) with $V_{\text{res}}$ the integration domain. But if we use the scattered field $\Psi_{\text{sca}}$ to get $P_{\text{abs}}$, because the scattered field $\mathbf{E}_{\text{sca}}$ is not always equal to the total field $\mathbf{E}_{\text{total}}$, an extra term $\iiint_{V_{\text{res}}} \text{Im}(A\epsilon_0\epsilon_{\text{rro}})\left(|\mathbf{E}_{\text{sca}} + \mathbf{E}_{\text{bg}}|^2 - |\mathbf{E}_{\text{sca}}|^2\right)d^3\mathbf{r}$ is needed. which is also necessary when obtaining $P_{\text{ext}}$. After simplification, the expression for $P_{\text{abs}}$ is

$$P_{\text{abs}} = \iiint_{V_{\text{res}}} Im\left[A\epsilon_0\epsilon_{r\infty}|\mathbf{E}_{\text{sca}}+\mathbf{E}_{\text{bg}}|^2 + \frac{|\Omega|^2}{2A\epsilon_0\epsilon_{r\infty}}\left(|\mathbf{P}_{1,\text{sca}}|^2+|\mathbf{P}_{2,\text{sca}}|^2\right) - \frac{\omega\Omega^*}{2A\epsilon_0\epsilon_{r\infty}}\left(|\mathbf{P}_{1,\text{sca}}|^2-|\mathbf{P}_{2,\text{sca}}|^2\right)\right]d^3\mathbf{r} \quad (30)$$

where $\mathbf{E}_{\text{sca}}$, $\mathbf{P}_{1,\text{sca}}$ and $\mathbf{P}_{2,\text{sca}}$ can be expanded in the form of Eq. (25). For material dispersion case of multiple partial fractions, $P_{\text{abs}}$ becomes

$$\begin{aligned}P_{\text{abs}} = \sum_{j=1}^{N}\iiint_{V_{\text{res}}} Im\bigg[&A_j\epsilon_0\epsilon_{r\infty}|\mathbf{E}_{\text{sca}}+\mathbf{E}_{\text{bg}}|^2 + \frac{|\Omega_j|^2}{2A_j\epsilon_0\epsilon_{r\infty}}\left(|\mathbf{P}_{1j,\text{sca}}|^2+|\mathbf{P}_{2j,\text{sca}}|^2\right)\\ &-\frac{\omega\Omega_j^*}{2A_j\epsilon_0\epsilon_{r\infty}}\left(|\mathbf{P}_{1j,\text{sca}}|^2-|\mathbf{P}_{2j,\text{sca}}|^2\right)\bigg]d^3\mathbf{r}.\end{aligned} \quad (31)$$

The expression of $P_{\text{ext}}$ is

$$\begin{aligned}P_{\text{ext}} = \frac{1}{2}\iiint_{V_{\text{res}}} Im\Big\{&\left[\omega\epsilon_0(\epsilon_{r\infty}-\epsilon_{rbg}^*)-(A-A^*)\epsilon_0\epsilon_{r\infty}\right]\mathbf{E}_{\text{sca}}\cdot\mathbf{E}_{\text{bg}}^* + \Omega\mathbf{P}_{1,\text{sca}}\cdot\mathbf{E}_{\text{bg}}^*\\ &-\Omega^*\mathbf{P}_{2\text{sca}}\cdot\mathbf{E}_{\text{bg}}^*\Big\}d^3\mathbf{r} + \iiint_{V_{\text{res}}} Im(A)\epsilon_0\epsilon_{r\infty}\left(|\mathbf{E}_{\text{sca}}+\mathbf{E}_{\text{bg}}|^2-|\mathbf{E}_{\text{sca}}|^2\right)d^3\mathbf{r}\end{aligned} \quad (32)$$

And for material dispersion case of multiple partial fractions, $P_{\text{ext}}$ becomes

$$\begin{aligned}P_{\text{ext}} = \frac{1}{2}\iiint_{V_{\text{res}}} Im\Bigg\{&\left[\omega\epsilon_0(\epsilon_{r\infty}-\epsilon_{rbg}^*)-\sum_{j=1}^{N}(A_j-A_j^*)\epsilon_0\epsilon_{r\infty}\right]\mathbf{E}_{\text{sca}}\cdot\mathbf{E}_{\text{bg}}^* + \sum_{j=1}^{N}\Omega_j\mathbf{P}_{1j,\text{sca}}\cdot\mathbf{E}_{\text{bg}}^*\\ &-\sum_{j=1}^{N}\Omega_j^*\mathbf{P}_{2j,\text{sca}}\cdot\mathbf{E}_{\text{bg}}^*\Bigg\}d^3\mathbf{r} + \sum_{j=1}^{N}\iiint_{V_{\text{res}}} Im(A_j)\epsilon_0\epsilon_{r\infty}\left(|\mathbf{E}_{\text{sca}}+\mathbf{E}_{\text{bg}}|^2-|\mathbf{E}_{\text{sca}}|^2\right)d^3\mathbf{r}.\end{aligned} \quad (33)$$

It can be proven that Eqs. (30) ~ (33) are equivalent to the general definition of absorption power and extinction power [8,43].

Consequently, the absorption cross section $\sigma_{\text{abs}}$ and extinction cross section $\sigma_{\text{ext}}$ can expressed as

$$\sigma_{\text{abs}} = \frac{P_{\text{abs}}}{I_{\text{inc}}}, \sigma_{\text{ext}} = \frac{P_{\text{ext}}}{I_{\text{inc}}} \quad (34)$$

where $I_{\text{inc}}$ is the incident light intensity. Finally, the scattering cross section is $\sigma_{\text{sca}} = \sigma_{\text{ext}} - \sigma_{\text{abs}}$.

### 2.4. Diffraction efficiency

For the diffraction problem of periodic structure, to obtain the diffraction efficiencies of reflection/transmission orders, it is necessary to reconstruct the total field distribution at a certain plane parallel to the grating plane in the reflection/transmission media. Due to the completeness assumption of QNMs and PML modes, it is also rigorous to reconstruct the field "outside" the resonator. For simplicity, we only investigate the 2D periodic structures with rectangular unit cell with periodic directions along x and y directions and lattice constants $p_x$ and $p_y$, and we also assume that the incident wave is from the top semi-infinite space (denoted as "+1") with real constant permittivity $\epsilon_r^{(+1)}$, and the transmission media is the bottom semi-infinite space (denoted as "-1") with permittivity $\epsilon_r^{(-1)}$, as is shown in Fig. 4. The wavenumbers are $k^{(\pm 1)} = 2\pi\sqrt{\epsilon_r^{(\pm 1)}}/\lambda$ with $\lambda$ the wavelength in vacuum. The incident angle is $\theta$ and the azimuth angle is $\varphi$. **p** and **s** denote the components of incident field oscillating in and perpendicular to the incident plane, respectively. Here a linear monochromatic plane wave is assumed with oscillating plane **u** in the **p-s** plane which has a $\psi$ angle respect to **p**.

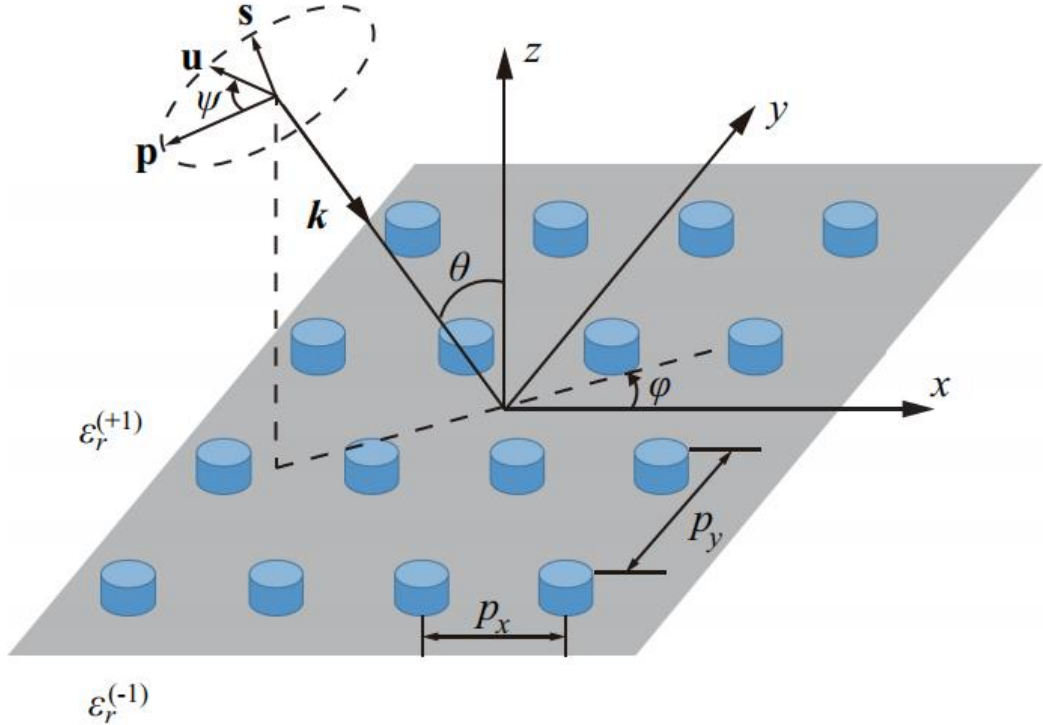

Fig. 4 Schematic of a plane wave incident on a periodic structure.

For two planes $z = z_0^{(+1)}$ and $z = z_0^{(-1)}$ in the reflection space and transmission space, respectively, on one hand, we can obtain the diffracted field $\Psi_{k_b,\text{dif}}\left(z = z_0^{(\pm 1)}\right)$ according to the QNMEM, and thus the total field can be expressed as

$$\Psi_{k_b,\text{ total}}\left(z = z_0^{(\pm 1)}\right) = \Psi_{k_b,\text{ dif}}\left(z = z_0^{(\pm 1)}\right) + \Psi_{k_b,\text{gg}}\left(z = z_0^{(\pm 1)}\right). \qquad (35)$$

On the other hand, according to the Floquet-Bloch Theorem, $\Psi_{\mathbf{k}_b,\text{total}}\left(z=z_0^{(\pm 1)}\right)$ can be expanded to Rayleigh series as [44]

$$\Psi_{\mathbf{k}_b,\text{total}}\left(z=z_0^{(+1)}\right) = \Psi_{\text{inc}}\exp\left[i\left(\alpha_0 x + \beta_0 y - \gamma_{00}^{(+1)} z_0^{(+1)}\right)\right] \\ + \sum_{q,v}\Psi_{R,qv}\exp\left[i\left(\alpha_q x + \beta_v y + \gamma_{qv}^{(+1)} z_0^{(+1)}\right)\right] \quad (36)$$

$$\Psi_{\mathbf{k}_b,\text{total}}\left(z=z_0^{(-1)}\right) = \sum_{q,v}\Psi_{T,qv}\exp\left[i\left(\alpha_q x + \beta_v y - \gamma_{qv}^{(-1)} z_0^{(-1)}\right)\right] \quad (37)$$

where $\Psi_{\text{inc}}$, $\Psi_{R,qv}$ and $\Psi_{T,qv}$ are the incident field, $(q, v)$ order reflected field and $(q, v)$ order transmitted field; and $\alpha_0$, $\beta_0$ and $-\gamma_{00}^{(+1)}$ are $x$, $y$ and $z$ projection of the incident wavevector, and can be written as

$$\alpha_0 = k^{(+1)}\sin\theta\cos\varphi,\ \beta_0 = k^{(+1)}\sin\theta\sin\varphi,\ \gamma_{00}^{(+1)} = k^{(+1)}\cos\theta \quad (38)$$

where the Bloch vector $\mathbf{k}_b$ corresponds to the in-plane wave vector components of incident wave due to the phase matching, i.e., $\mathbf{k}_b = \hat{\mathbf{x}}\alpha_0 + \hat{\mathbf{y}}\beta_0$ with $\hat{\mathbf{x}}$ and $\hat{\mathbf{y}}$ unit vectors. $\alpha_q$, $\beta_v$ an $\gamma_{qv}^{(\pm 1)}$ are the projection components for the diffracted orders, and can be expressed as

$$\alpha_q = \alpha_0 + qK_x,\ \beta_v = \beta_0 + vK_y, \quad (39)$$

where $K_x = 2\pi/p_x$ and $K_y = 2\pi/p_y$, and $\gamma_{qv}^{(\pm 1)}$ is

$$\gamma_{qv}^{(\pm 1)} = \sqrt{k^{(\pm 1)2} - \alpha_q^2 - \beta_v^2}. \quad (40)$$

To guarantee $\gamma_{qv}^{(+1)}$ and $\gamma_{qv}^{(-1)}$ in Eqs. (36) and (37) correspond to propagation or decaying planes waves in $+z$ and $-z$ directions, respectively, there should be

$$\text{Re}\left(\gamma_{qv}^{(\pm 1)}\right) + \text{Im}\left(\gamma_{qv}^{(\pm 1)}\right) > 0 \quad (41)$$

Owing to the orthogonality of the Fourier series, $\Psi_{R,qv}$ and $\Psi_{T,qv}$ can be retrieved as

$$\Psi_{R,qv} = \frac{1}{p_x p_y}\int_{-\frac{p_y}{2}}^{\frac{p_y}{2}}\int_{-\frac{p_x}{2}}^{\frac{p_x}{2}}\left\{\Psi_{\mathbf{k}_b,\text{total}}\left(z=z_0^{(+1)}\right) - \Psi_{\text{inc}}\exp\left[i\left(\alpha_0 x + \beta_0 y - \gamma_{00}^{(+1)} z_0^{(+1)}\right)\right]\right\} \\ \cdot \exp\left[-i\left(\alpha_q x + \beta_v y + \gamma_{qv}^{(+1)} z_0^{(+1)}\right)\right]dx\,dy, \quad (42)$$

$$\Psi_{T,qv} = \frac{1}{p_x p_y}\int_{-\frac{p_y}{2}}^{\frac{p_y}{2}}\int_{-\frac{p_x}{2}}^{\frac{p_x}{2}}\Psi_{\mathbf{k}_b,\text{total}}\left(z=z_0^{(-1)}\right)\cdot\exp\left[-i\left(\alpha_q x + \beta_v y - \gamma_{qv}^{(-1)} z_0^{(+1)}\right)\right]dx\,dy. (43)$$

Substituting Eqs. (27) and (35) into Eqs. (42) and (43), it can be inferred that the diffraction order field can be expanded into the superposition of integration of QNMs term by term and an extra term of integration related with the background field.

Having obtained the complex amplitude of each propagation order, the diffraction efficiency can be calculated correspondingly. It can be expressed in the form of electric field as

$$\eta_{R,qv} = \frac{\gamma_{qv}^{(+1)}|\mathbf{E}_{R,qv}|^2}{\gamma_{00}^{(+1)}|\mathbf{E}_{\text{inc}}|^2}, \eta_{T,qv} = \frac{\gamma_{qv}^{(-1)}|\mathbf{E}_{T,qv}|^2}{\gamma_{00}^{(+1)}|\mathbf{E}_{\text{inc}}|^2} \qquad (44)$$

and they can also be expressed in the form of magnetic field as

$$\eta_{R,qv} = \frac{\gamma_{qv}^{(+1)}|\mathbf{H}_{R,qv}|^2}{\gamma_{00}^{(+1)}|\mathbf{H}_{\text{inc}}|^2}, \eta_{T,qv} = \frac{\epsilon_r^{(+1)}\gamma_{qv}^{(-1)}|\mathbf{H}_{T,qv}|^2}{\epsilon_r^{(-1)}\gamma_{00}^{(-1)}|\mathbf{H}_{\text{inc}}|^2} \qquad (45)$$

If the constituting material is lossless, the summation of all diffraction efficiencies should equal 1. Otherwise, the absorption $Abs$ can be obtained by

$$Abs = 1 - \sum_{(q,v) \in U^{(+1)}} \eta_{R,qv} - \sum_{(q,v) \in U^{(-1)}} \eta_{T,qv}. \qquad (46)$$

## 3. Numerical demos

In this section, we will use two demos to show the performance of the built QNMEM, one is the Mie scattering of a metal nanosphere, and the other is the diffraction efficiencies of a 1D subwavelength metal-dielectric-metal (MDM) grating.

### 3.1. Scattering of a metal nanosphere

We here study the scattering of a metal nanosphere embedded in air. The metal nanosphere has a radius of $r_0$ =30nm, and its permittivity can be modeled by a single pair of partial fraction with parameters $\epsilon_{r\infty} = 1.5, A = (-3 \times 10^{17} + i2 \times 10^{14})$ rad/s. Its scattering/absorption/extinction cross section can also be predicted semi-analytically by Mie theory, which can be used as reference to verify the results reconstructed by QNMEM here, also the poles can also be predicated by Mie theory which can be used to distinguish QNMs and PML modes here. It is noteworthy that the eigenfrequency could be trapped around the material poles and zeros which correspond to longitude electromagnetic modes and are usually nonphysical, and should be detoured during the computation. Besides, for metal material, mode aggregation could appear around SPP frequency of metal-dielectric interface where $\epsilon(\widetilde{\omega}_m) + \epsilon_{\text{rbg}} = 0$. In this case, the pole and zero of the material dispersion are $\omega_{\text{pole}} = (\pm 1.5 \times 10^{14} - i5 \times 10^{13})$ rad/s and $\omega_{\text{zero}} = (\pm 9.486 \times 10^{15} - i2.500 \times 10^{14})$ rad/s, respectively, and the SPP frequency is $\omega_{\text{SPP}} = (\pm 7.349 \times 10^{15} - i1.700 \times 10^{14})$ rad/s. In In the simulation, the smallest mesh is $r_0/5$ for the nanosphere domain, and is $r_0$ for the air domain and the PML.

The concerned frequency range is $[3.319 \times 10^{15}\text{rad/s}, 1.256 \times 10^{16}\text{rad/s}]$ (or wave-length range of [150 nm, 600 nm]). We can separate the contribution of each mode according to one variant of Eqs. (32) and (33) (see Supplement 4)

$$\sigma_{\text{ext}} = \sum_{m=1}^{\infty} \frac{\omega}{2I_{\text{inc}}} \iiint_{V_{\text{res}}} \text{Im}\{\epsilon_0[(\epsilon_r(\widetilde{\omega}_m) - \epsilon_{\text{rbg}}^*)]\Lambda_m(\omega)\widetilde{\mathbf{E}}_m \cdot \mathbf{E}_{\text{bg}}^*\}d^3\mathbf{r}. \quad (47)$$

The eigenmodes can be ranked in the descending order according to their average contribution $<\sigma_{\text{ext},m}>_{\text{avg}}$ in the interested range shown in Fig. 5, which indicates that the dominant modes are the electric dipole mode at $(5.535 \times 10^{15} - i3.747 \times 10^{14})$ rad/s and electric quadrupole mode at $(6.511 \times 10^{15} - i1.453 \times 10^{14})$ rad/s. Except for part of them, most PML modes contribute rarely to the reconstruction.

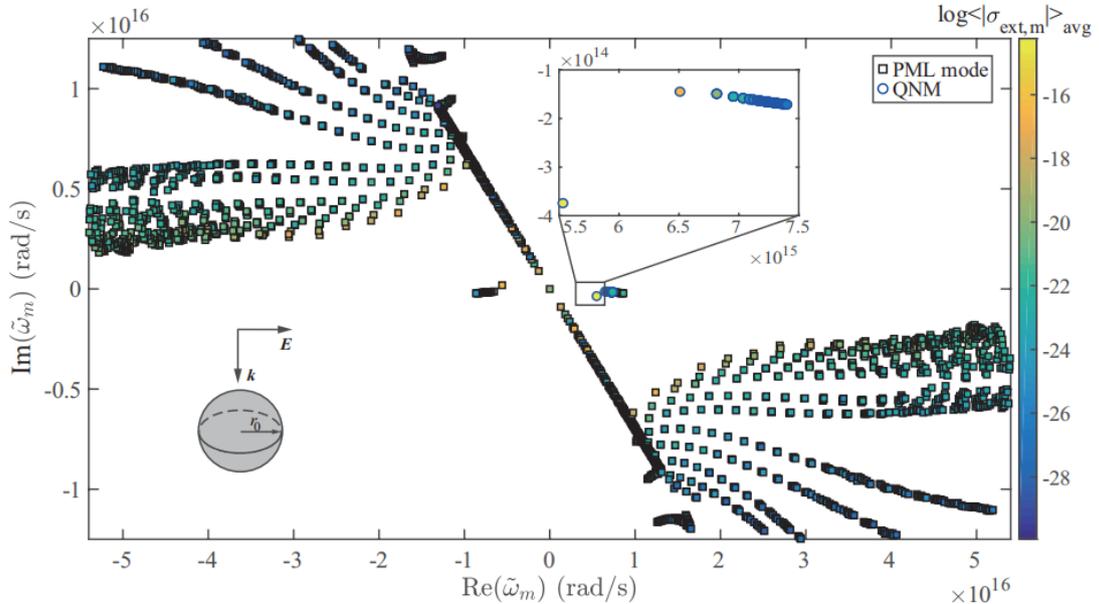

Fig. 5. Distribution of eigenfrequencies of the metal nanosphere with a radius of 30 nm. The circles denote the QNMs, squares denote PML modes, and filled color denotes the magnitude of the average contribution $<\sigma_{\text{ext},m}>_{\text{avg}}$ of each eigenmode. The enlarged inset shows the distribution of QNMs in the studied range.

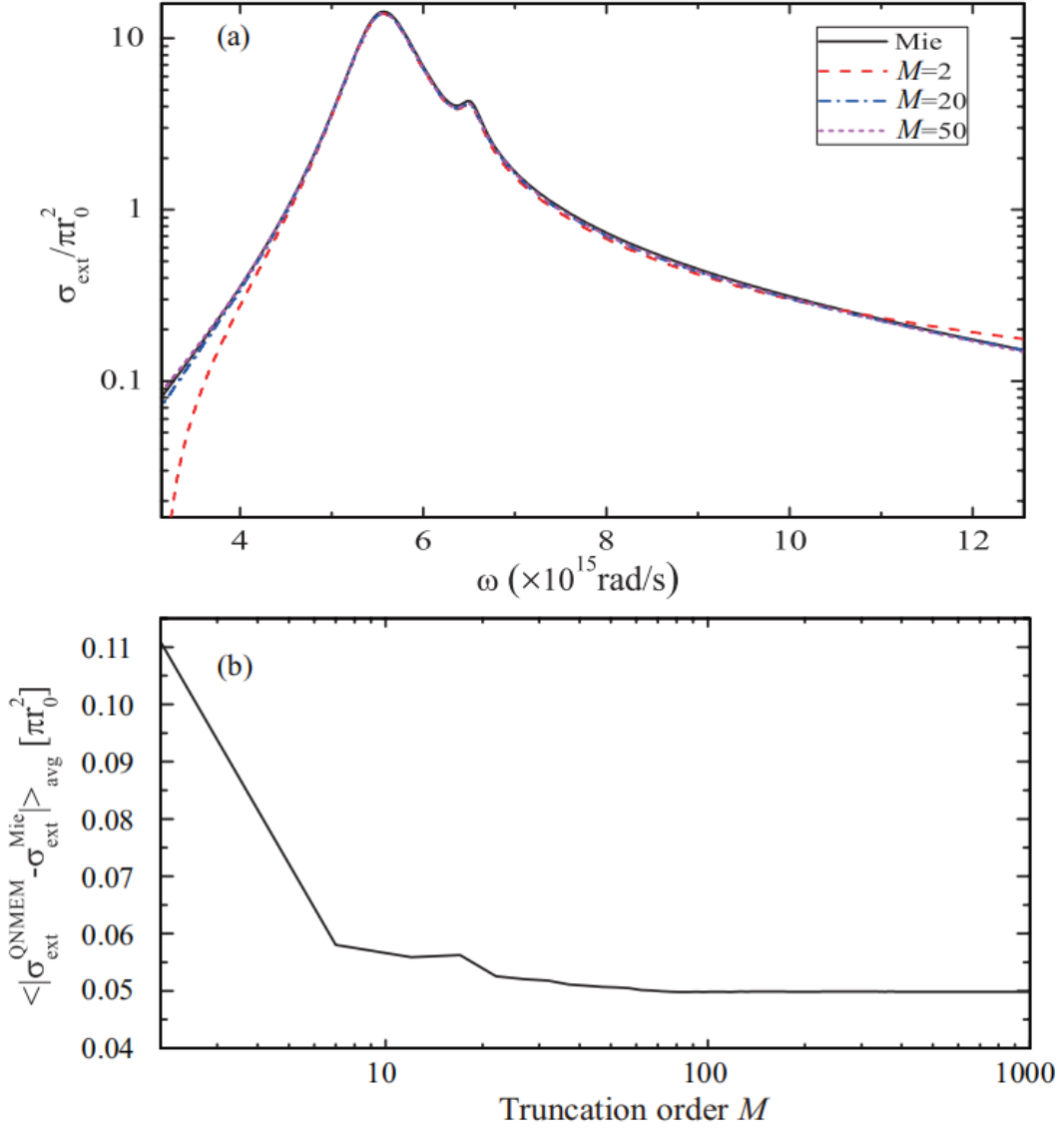

Fig. 6 (a) Extinction cross section by Mie theory and reconstructed by $M$=2, 20, 50 eigenmodes. The eigenmodes are ranked in the descending order according to their average contributions $<\sigma_{ext,m}>_{avg}$. For $M$=2, the selected eigenmodes are the electric dipole mode and the electric quadrupole mode. (b) Convergence curve for QNMEM. With the increasing of $M$, the average error $<|\sigma_{ext}^{QNMEM} - \sigma_{ext}^{Mie}|>_{avg}$ converges to a constant about 0.05. Note that the extinction cross section here is in the unit of $\pi r_0^2$.

Real computation needs to truncate the eigenmode order $M$, or the number of eigenmodes involved in the reconstruction. Figure 6(a) shows the extinction cross section spectrum reconstructed with $M$=2, 20, 50, which implies that even only 2 modes are enough to reconstruct the decent result. With more eigenmodes considered in the reconstruction, the extinction cross section gradually approximates the Mie prediction.

Figure 6(b) shows the convergence of the average extinction cross section spectrum error $< |\sigma_{\text{ext}}^{\text{QNMEM}} - \sigma_{\text{ext}}^{\text{Mie}}| >_{\text{avg}}$ where $\sigma_{\text{ext}}^{\text{QNMEM}}$ is the extinction cross section by the QNMEM while $\sigma_{\text{ext}}^{\text{Mie}}$ is by the Mie theory. With the increasing of truncation order $M$, the average error converges to a steady value. Two factors stop the ultimate error from continuing decreasing: (1) high order PML modes' contribution is insignificant and make little effort on decreasing the average error; (2) the accuracy of numerical simulation is limited by the size of mesh [3] .

The computation time of eigenmodes and reconstruction in QNMEM is related with the truncation order $M$, and the first part is dominant [3] . When $M$ is small, QNMEM has advantage over traditional methods like the frequency domain FEM, at the cost of some accuracy lost. Besides, eigenmodes with eigenfrequencies closer to the incident field are excited stronger and contribute more to the total response, making the physical meaning of the QNMEM intuitive.

### 3.2. Diffraction efficiency of MDM grating

Here we study the diffraction of the 1D MDM subwavelength grating under TM polarization (or p polarization) plane wave in collinear mounting. A unit cell of the 1D MDM grating is shown in Fig. 7 with period $p$ = 350 nm. The substrate is metal which can shield the transmission; the middle layer is an ultra thin dielectric gap layer with a thickness of $t_{\text{di}}$ = 15 nm; and the top layer is a metal grid array with a width of $w$= 250 nm and a thickness of $t_{\text{me}}$ = 20 nm. The metal is gold, the relative permittivity of which can be described by Drude model as [45, 46] $\epsilon_{r,\text{Au}} = 1 - \frac{\omega_p}{\omega^2 + i\gamma\omega}$ with $\omega_p =$ 1.32 × 10$^{16}$ rad/s and $\gamma = 1.2 \times 10^{14}$ /s, and we can convert it the form of partial fraction according to Tab. 2. The dielectric is SiO2 with a relative permittivity of $\epsilon_{r,\text{SiO}_2} = 2.25$. The incident media is air with $\epsilon_{r,\text{Air}} = 1$. According to Sec. 1.2.4, there is $k_{\text{b}} = k_x = \alpha_0 = \frac{2\pi}{\lambda}\sin(\theta)$, where $\lambda$ is the vacuum wavelength. In this case, $k_{\text{b}}$ is fixed, except for $k_{\text{b}} = 0$ ($\theta = 0$), for general cases when $k_{\text{b}} \neq 0$, the incident angle for different wavelength is different.

For normal incidence ($k_{\text{b}} = 0$), in the wavelength range of [600nm,2500nm], the

1D MDM grating satisfy the subwavelength condition $(\lambda > \sqrt{\epsilon_{r,\text{Air}}}p(1 + \sin(\theta)))$ and has only the $0^{\text{th}}$ reflected order. The absorption is thus $Abs = 1 - R_0$ where $R_0$ is the diffraction efficiency of $0^{\text{th}}$ reflected order.

Figure 8(a) compare the $R_0$ spectrum by different methods, i.e., the Fourier modal method (FMM), the frequency domain finite element method(FD FEM) and the QNMEM. In this study, the FMM is implemented by freeware Reticolo [47] with 601 truncation Fourier harmonics; the FD FEM is implemented by frequency domain solver of the commercial software COMSOL Multiphysics with periodic ports. In the QNMEM, the reconstruction plane is 30 nm above the grating, and the background field can be obtained analytically by a transfer matrix method [48] . The FD FEM and QNMEM use the same meshes with a largest size of 10 nm in the metal and dielectric domain while 100 nm in the air domain, and an equivalent mesh size of 200 nm in the PML. We compare the performance of the reconstruction of QNMEM with different $M$. Note that the eigenmodes are ranked in the descending order of their average spectra contribution $<|\Lambda_{\mathbf{k}_b,m}|>_{\text{avg}}$ .

From Fig. 8(a), we can conclude that the results by FMM and FD FEM are consistent basically, and the spectrum obtained by the FMM is only a little blue-shift relative to that by FD FEM, thus we can take the result from FD FEM as "accurate" reference. There are two apparent dips in $R_0$ spectrum, i.e., $R_0(1860 \text{ m}) = 0.12\%$

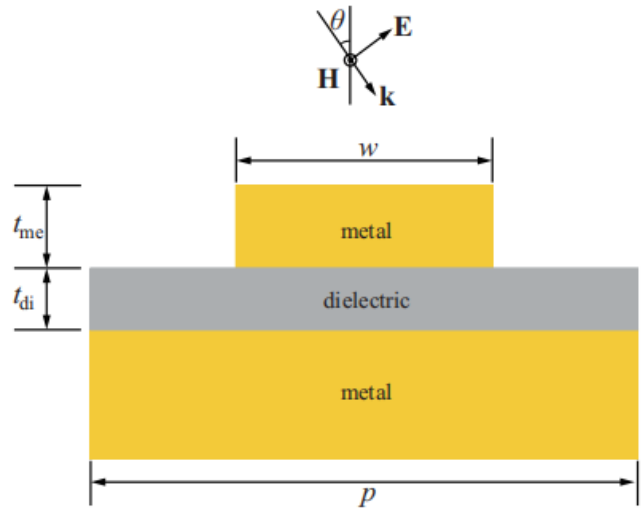

Fig. 7 Schematic of 1D MDM grating.

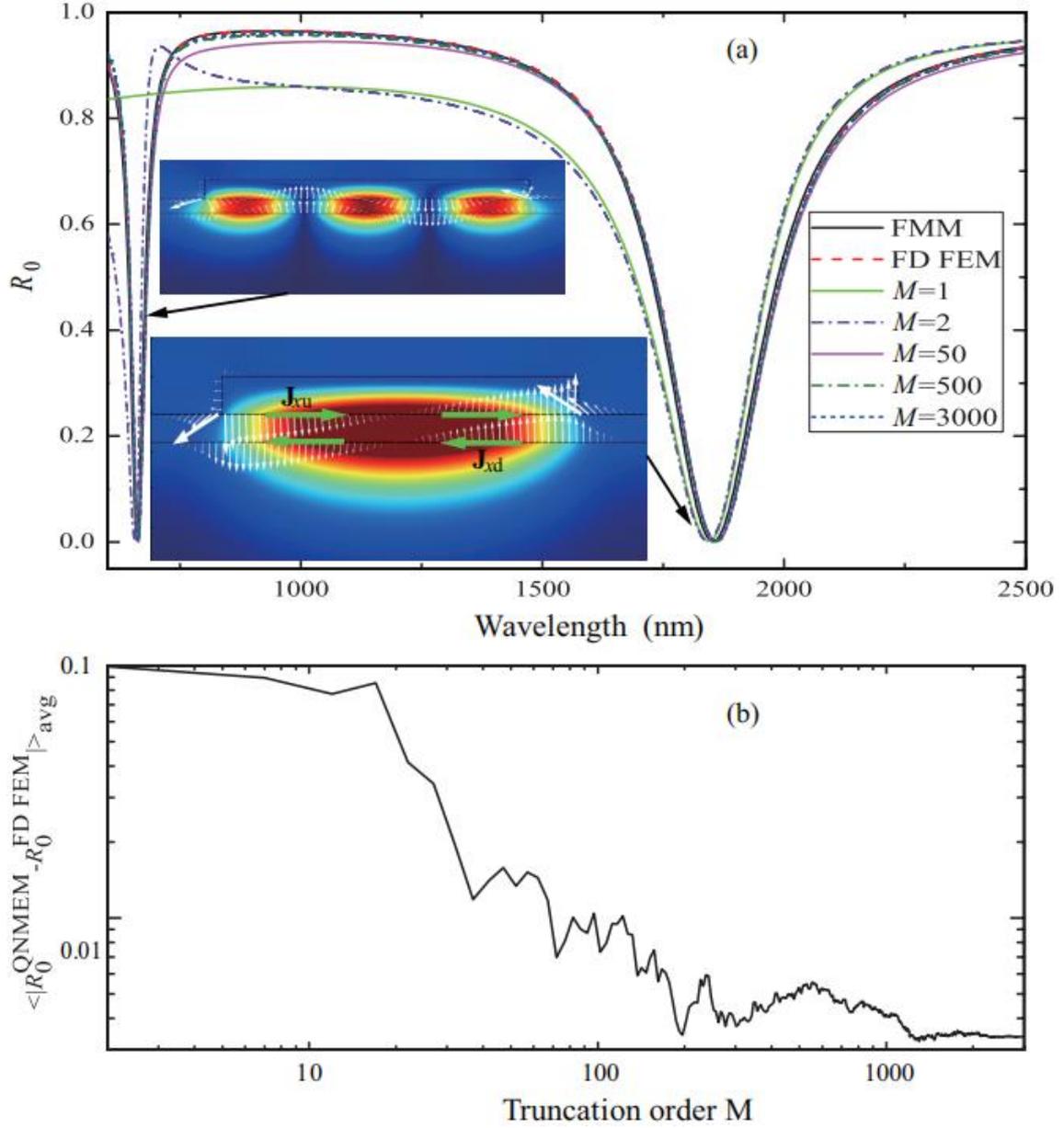

Fig. 8. (a) In the case of $k_b = 0$, the diffraction efficiency of $0^{th}$ reflected order $R_0$ by FMM, FD FEM and QNMEM with different $M$. Two reflection dips appear around 1860 nm and 664 nm, and the insets show the distribution of $|H_y|$ (color map) and current density (white arrows) of 2 QNMs around them, also the bold green arrows in the lower inset indicate the in-plane current density $J_{xu}$ and $J_{xd}$ along the upper and lower metal-dielectric interface, respectively. (b) The converge of $R_0$ by QNMEM. With the increasing of $M$, the average spectrum error $<|R_0^{QNMEM} - R_0^{FD\ FEM}|>_{avg}$ decreases stable and gradually approximates about $3.4\times10^{-3}$.

and $R_0(664nm) = 0.29\%$, which means the 1D MDM grating has nearly perfect absorption around these two wavelength. With the increasing of $M$, the reconstructed

$R0$ by the QNMEM gradually approximates that by the FD FEM globally. It is notable that even when $M = 1$, the QNMEM can reconstruct the $R_0$ spectrum around 1860 nm well, and when $M = 2$, both the two dips can be reconstructed well. Note that the precise locations of the two dips in $R_0$ spectrum deviate the real part of the eigenfrequencies somewhat, which is caused by the complex Fano interference between modes and background field [49, 50].

Figure 8(a) also show the distribution of |Hy| and the current density of two QNMs around the two dips. The magnetic field of the two QNMs are highly localized in the gap area, and the in-plane current component in the upper metal-dielectric interface $J_{xu}$ and in the lower metal-dielectric interface $J_{xd}$ are out of phase, forming loop current, which indicates that these two QNMs are the so-called magnetic resonance. Besides, the magnetic field pattern is very similar to the standing-wave in the F-P resonator, which is in fact built by the constructive interference of gap surface plamson [51, 52]. The eigen wavelength of the fundamental mode is $\tilde{\lambda}_1 = 2\pi c/\tilde{\omega}_1 = (1856.12 + 128.27i)$ nm, and its magnetic field |Hy| has one antinode (the electric field |E| has a node correspondingly), which is very similar to the fundamental mode of the F-P resonator [48, 51, 52], that's why we call it fundamental mode here, and we denote its interference order as $l = 1$. Similarly, the other QNM at $\tilde{\lambda}_2 = (663.49 + 17.22i)$ nm corresponds to $l = 3$. Please not confound the denotation of interference order and eigenmode order, the former is only applicable to the situation in this case.

Figure 8(b) shows the convergence curve of $R_0$ by QNMEM with the result by FD FEM as reference. With the increasing of $M$, the average spectra error $<|R_0^{\text{QNMEM}} - R_0^{\text{FD FEM}}|>_{\text{avg}}$ decreases gradually and converges to a stable value, which shows similar characteristics as those in Fig. 6(b) for the scattering problem and indicates that the QNMEM can be used as a rigorous method to reconstruct the diffraction efficiency spectra. Meanwhile, in the case for fast reconstruction the profile with reasonable accuracy and vivid physical picture of resonant process, only retaining the several leading QNMs are enough.

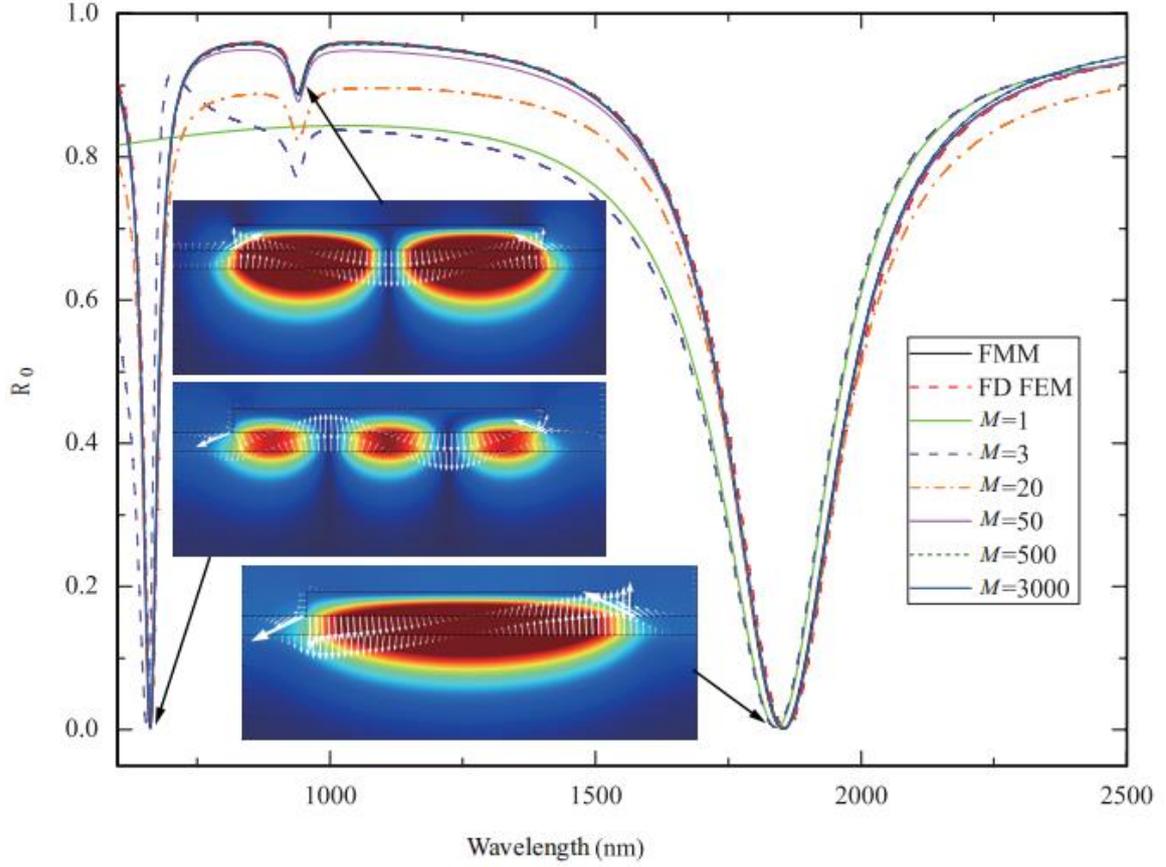

Fig. 9. For the case of $k_b = 0.1\pi/p$, the diffraction efficiency of $0^{th}$ reflected order $R_0$ by FMM, FD FEM and QNMEM with different $M$. Three dips at about 1859 nm, 942 nm and 664 nm. The insets show the distribution of $|H_y|$ (color map) and current density of 3 QNMs around these dips (white arrows).

In the inclined incidence of $k_b = 0.1\pi/p$, the grating still satisfies the subwavelength condition in [600nm,2500nm], and only the $R_0$ is concerned on. Similar to Fig. 8(a), Fig. 9 compares $R_0$ obtained by different methods. Only 3 QNMs are enough to reconstruct the rough profile of the $R_0$ spectrum and 3 reflection dips around 1859 nm ($R_0$ = 0. 15%), 942 nm ($R_0$= 88.89%) and 664 nm ($R_0$= 0.42%), which means two near perfect absorption peaks and a weak absorption peak. With the increasing of $M$, the result by QNMEM gradually approximates those be the FD FEM and the FMM.

Fig. 9 also shows the distribution of magnetic field $|H_y|$ and current of 3 QNMs corresponding to the 3 reflection dips, which show similar characteristics with those in Fig. 8(a). Among them, the QNM with $\tilde{\lambda}_1 = 2\pi c/\tilde{\omega}_1 = (1851.16 + 128.32i)$ nm.

can be denoted as $l=1$, the one with $\tilde{\lambda}_3 = 2\pi c/\tilde{\omega}_3 = (940.49 + 17.23i)$ nm can be denoted as $l = 2$, and the one with $\tilde{\lambda}_2 = 2\pi c/\tilde{\omega}_2 = (662.32 + 16.87i)$ nm can be denoted as $l = 3$.

The real part of eigen wavelengths of these 3 QNMs approximately satisfy $1:\frac{1}{2}:\frac{1}{3}$, note that the notation order of QNM is based on their contribution while the notation order of interference order is based on the numbers of antinodes of $|H_y|$. Compared with the normal incidence condition, two features are worth pointing out, the first is that the eigen wavelength of $l=1$ and $l=3$ order only deviates a little in the inclined incidence which indicates that these modes are localized resonance modes rather than collective (prorogation) resonance modes; the second is that the effect of the $l = 2$ order only appears in the inclined incidence. In fact, a QNM corresponding to $l=2$ with eigen wavelength $\tilde{\lambda}_3 = (941.94 + 16.95i)$ nm can be obtained when $k_b = \alpha_0 = 0$, but because the electric field of both the incident field and $0^{th}$ reflected order only has only the x component. Due to the symmetry of the structure, $E_x$ of QNMs with even $l$ is an anti-symmetric to the axis of symmetry, while the background filed is symmetric, thus the excitation coefficients of these modes are 0 according to Eqs. (28), (35) and (43), and they have no contributions to the ultimate output spectra and no resonance features can be observed around these wavelengths in $R_0$ spectrum. In fact, if the constituting material is lossless, these QNMs are just a symmetry-protected BIC states with accidentally real eigenfrequencies [53]. Only when $k_b \neq 0$, this symmetry is broken, and the contribution of these modes would appear gradually.

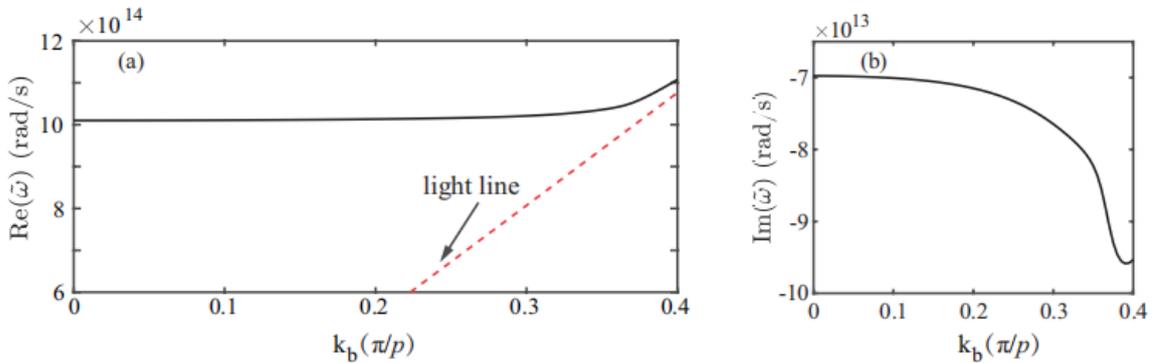

Fig. 10 (a) Dispersion relation for the fundamental modes of the 1D MDM grating. (b) Change of the imaginary part of the eigenfrequency with $k_b$.

From Fig. 8(a) and 1.9, we can find that with only one fundamental mode retained, we can reconstructed the $R_0$ around 1859 nm. The reflection dip and absorption peak around the fundamental wavelength are more stable with the change of incident angle [54, 55]. Besides, the fundamental wavelength is also further away from the Rayleigh anomalies and propagation surface plasmon resonance, so usually the fundamental mode is more concerned. Here we will study the change of fundamental mode eigenfrequency with $k_b$, i.e. the dispersion relation $\text{Re}(\widetilde{\omega}) - k_b$, as is shown in Fig. 10(a). It is shown that $\text{Re}(\widetilde{\omega})$ keeps nearly a horizontal line up to $k_b \leqslant 0.35\pi/p$, or keeps nearly 1855 nm up to $\theta \leqslant 68.05°$, which verifies the localization of the fundamental mode of the MDM grating and insensitivity to the incident angle. In the meantime, Fig. 10(b) shows that $\text{Im}(\widetilde{\omega})$ varies slowly when $k_b \leqslant 0.2\pi/p$ ($\theta \leqslant 32°$ at 1855 nm), indicating that MDM grating should have wide-angle anti-reflection property.

To check the conclusions above, Fig. 11 compares the reconstructed $R_0$ by QNMEM with $M = 1$ and the accurate $R_0$ by FD FEM at different $k_b$, which are consistent basically around the resonant frequency. As a consequence, for initial design, we can use a few QNMs and even only one QNM to reconstruct the approximate $R_0$ spectrum, and find a good initial value in a high dimensional parameter space, and then refine the structure around the initial values by other rigorous method like QNMEM with large $M$ or just the FD FEM, or the hybrid-optimization method. Besides, Fig. 11 also verify the insensitivity of the resonant wavelength of the subwavelength MDM grating, and the resonant reflectivity only increase a little (< 10.8%) in a large incident angle range from $0°$ to $80°$. Thus only considering the normal incidence is necessary in the initial design of this kind of perfect absorber based on localized mode. Note that $R_0$ in 1.11(a) deviates significantly with that by the FD FEM at large incident angle in the high frequency due to the effect of collective resonances.

The Drude model dispersion in the example above is quite simple, which is not always proper for real materials where dispersion models with more partial fractions are needed. Fitted from the experimental data of Johnson et. al. [56], the permittivity of gold can be modeled by two pairs of partial fractions [7] with $\epsilon_{r\infty} = 1$, $A_1 = (-2.6291492 \times 10^{17} + i1.3032853 \times 10^{15})$rad/s, $\Omega_1 = (3.1528585 \times 10^{14} - i5.0113345 \times 10^{13})$rad/s, $A_2 = (-2.0151265 \times 10^{15} + i1.1833388 \times 10^{16})$rad/s and $\Omega_2 = (3.7903321 \times 10^{15} - i1.6977449 \times 10^{15})$ rad/s. The refraction index of SiO2 can be described by Sellmier formula as [57, 58] $n_{SiO_2}^2 - 1 = \frac{1.144606\lambda^2}{\lambda^2 - 0.08774721^2} + \frac{7.504816\lambda^2}{\lambda^2 - 490.4066^2}$ where the wavelength λ is in the unit of μm, and then it can be transformed into the form of partial fractions according to Tab. 1.

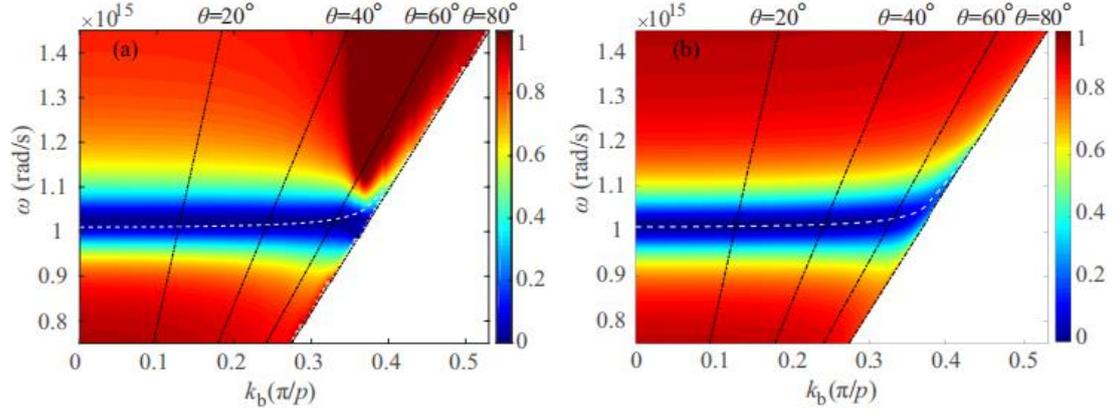

Fig. 11 Diffraction efficiency of 0th reflected order $R_0$ by (a) QNMEM ($M$=1) and (b) FD FEM. The white dash lines indicate the resonant frequencies of the fundamental mode, and the black dot dash lines show the locations of a series of incident angles.

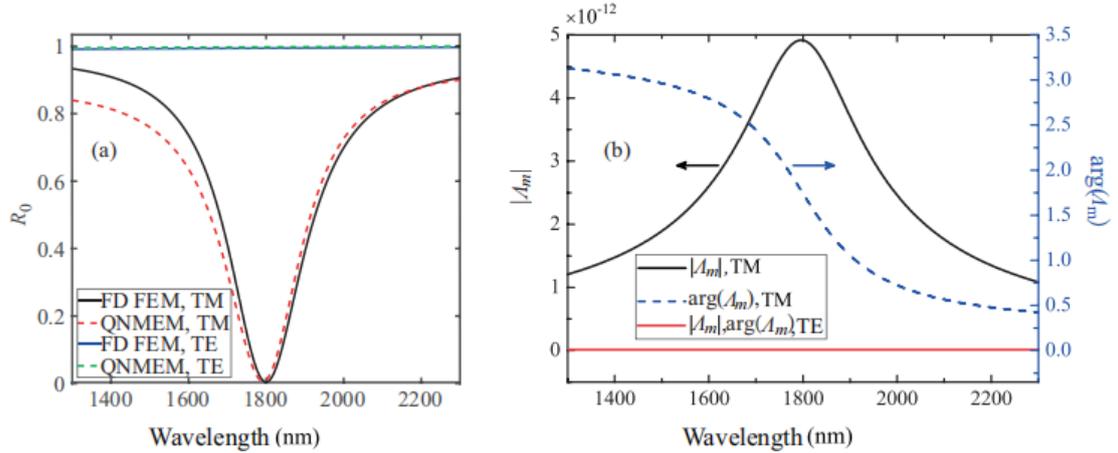

Fig. 12 (a) Comparison of the $R_0$ by QNMEM with M = 1 and by FD FEM in the case of complex dispersion model. (b) The magnitude and phase of the fundamental mode excitation coefficients under different polarization incidence.

Keeping the geometric parameters unchanged, Fig. 12(a) shows $R_0$ of the 1D

MDM grating around the fundamental wavelength by the QNMEM with $M$=1, which is also very close to the accurate result in TM (p) polarization incidence. Meanwhile, in TE (s) polarization, the structure is highly reflective without resonance feature because the incident field and background field only have components of $H_x$, $H_z$ and $E_y$, while the fundamental mode only has components $E_x$, $E_z$ and $H_y$, resulting in null excitation coefficients according to Eq. (28). Figure 12(b) shows the magnitude and phase of the excitation coefficients of the fundamental mode under TM polarization and TE polarization incidence, respectively, which clearly indicates that only the TM polarization incident field can be coupled to the structure effectively. Besides, the excitation coefficient under TM polarization is close to Lorentzian near the resonant wavelength [59], and phase undergoes a change of π across the resonant wavelength which are consistent with the typical resonant features.

## 4. Conclusions

In conclusion, we establish the QNMEM to evaluate the spectra of both nonperiodic and periodic nanoresonators. We first introduce auxiliary fields for the Partial-Fraction dispersion model to linearize the eigen equations, and build the solver for the augmented eigen equations to compute QNMs, and derive the QNMEM for scattering/diffraction problems. With the increase of truncation order $M$, the result reconstructed by the QNMEM converges to the accurate result. Around the resonant frequencies, retaining only a few leading QNMs can reconstruct the extinction cross section or diffraction efficiency decently. Although the convergence of the current QNMEM is still a little slow compared to mature methods such as FMM, FD FEM or FDTD, for structures with strong resonance, the QNMEM can reconstruct an approximate resonant spectrum fast, and reveal the underground physical nature at the same time, which are very useful for initial design. In the future, we will continue to develop faster QNMEM equipped with more information like zeros, R-zeros, T-zeros of the S matrix, and promote the application of the QNMEM to reality like the design of perfect absorbers.

# Acknowledgement

The author would appreciate the generous help from Philippe Lalanne and Wei Yan.

# Supplement 1  Unconjugated Lorentz reciprocity

We here derive the unconjugated Lorentz reciprocity after introducing the auxiliary fields, and two solutions of the Maxwell's equations $\widehat{H}\Psi_1 = \omega_1 \Psi_1 + S_1$ and $\widehat{H}\Psi_2 = \omega_2 \Psi_2 + S_2$ are needed, where $S_1$ and $S_2$ are sources of the two solutions, respectively. The sources can be either in near field or in the far field, and can be in the form of dipoles, magnetic dipoles, current source, plane waves and so on. Applying $\iiint_V d^3\mathbf{r} \Psi_2^T \widehat{D}$ to the two side of $\widehat{H}\Psi_1 = \omega_1 \Psi_1 + S_1$ gives

$$\iiint_V \Psi_2^T \cdot \widehat{D}\widehat{H}\Psi_1 d^3\mathbf{r} = \omega_1 \iiint_V \Psi_2^T \cdot \widehat{D}\Psi_1 d^3\mathbf{r} + \iiint_V \Psi_2^T \cdot \widehat{D}S_1 d^3\mathbf{r}, \qquad (S1\text{-}1)$$

where the integration domain $V$ is the whole PML mapped space including the PML. $\widehat{D}$ is a diagonal matrix, and for dispersion model with single pair of partial fraction it is

$$\widehat{D} = \mathrm{diag}\left[-\mu_0, \epsilon_0 \epsilon_{r\infty}, \frac{-\Omega}{A\epsilon_0 \epsilon_{r\infty}}, \frac{-\Omega^*}{A^* \epsilon_0 \epsilon_{r\infty}}\right]. \qquad (S1\text{-}2)$$

and for dispersion model with multiple pairs of partial fractions, it is

$$\widehat{D} = \mathrm{diag}\left[-\mu_0, \epsilon_0 \epsilon_{r\infty}, \frac{-\Omega_1}{A_1 \epsilon_0 \epsilon_{r\infty}}, \frac{-\Omega_1^*}{A_1^* \epsilon_0 \epsilon_{r\infty}}, \cdots, \frac{-\Omega_N}{A_N \epsilon_0 \epsilon_{r\infty}}, \frac{-\Omega_N^*}{A_N^* \epsilon_0 \epsilon_{r\infty}}\right]. \qquad (S1\text{-}3)$$

and for the dispersionless material, the auxiliary field is undefined, and there are $\Psi = [\mathbf{H}, \mathbf{E}]^T$ and $\widehat{D} = \mathrm{diag}\left[-\mu_0, \epsilon_0 \epsilon_{r\infty}\right]$. With simple algebra operations to the left-hand side of Eq. (S1-1), we can get

$$\iiint_V \Psi_2^T \cdot \widehat{D}\widehat{H}\Psi_1 d^3\mathbf{r}$$

$$= \iiint_V [\mathbf{H}_2, \mathbf{E}_2, \mathbf{P}_{12}, \mathbf{P}_{22}] \begin{bmatrix} 0 & i\nabla\times & 0 & 0 \\ i\nabla\times & -(A_j - A_j^*)\epsilon_0 \epsilon_{r\infty} & -\Omega & \Omega^* \\ 0 & -\Omega & \dfrac{-\Omega^2}{A\epsilon_0 \epsilon_{ro\infty}} & 0 \\ 0 & \Omega^* & 0 & \dfrac{\Omega^{*2}}{A^* \epsilon_0 \epsilon_{ro\infty}} \end{bmatrix} \begin{bmatrix} \mathbf{H}_1 \\ \mathbf{E}_1 \\ \mathbf{P}_{11} \\ \mathbf{P}_{21} \end{bmatrix} d^3\mathbf{r}$$

$$= \iiint_V [\mathbf{H}_1, \mathbf{E}_1, \mathbf{P}_{11}, \mathbf{P}_{21}] \begin{bmatrix} 0 & i\nabla \times & 0 & 0 \\ i\nabla \times & -(A_j - A_j^*)\epsilon_0 \epsilon_{ros} & -\Omega & \Omega^* \\ 0 & -\Omega & \frac{-\Omega^2}{A\epsilon_0 \epsilon_{r\infty}} & 0 \\ 0 & \Omega^* & 0 & \frac{\Omega^{*2}}{A^* \epsilon_0 \epsilon_{r\infty}} \end{bmatrix} \begin{bmatrix} \mathbf{H}_2 \\ \mathbf{E}_2 \\ \mathbf{P}_{12} \\ \mathbf{P}_{22} \end{bmatrix} d^3\mathbf{r}$$

$$+ i \iint_\Sigma (\mathbf{E}_1 \times \mathbf{H}_2 - \mathbf{E}_2 \times \mathbf{H}_1) \cdot d\mathbf{s}$$
$$= \iiint_V \mathbf{\Psi}_1^T \cdot \hat{\mathbf{D}} \hat{\mathbf{H}} \mathbf{\Psi}_2 d^3\mathbf{r} + i \iint_\Sigma (\mathbf{E}_1 \times \mathbf{H}_2 - \mathbf{E}_2 \times \mathbf{H}_1) \cdot d\mathbf{s}$$
$$= \omega_2 \iiint_V \mathbf{\Psi}_1^T \cdot \hat{\mathbf{D}} \mathbf{\Psi}_2 d^3\mathbf{r} + \iiint_V \mathbf{\Psi}_1^T \cdot \hat{\mathbf{D}} \mathbf{S}_2 d^3\mathbf{r} + i \iint_\Sigma (\mathbf{E}_1 \times \mathbf{H}_2 - \mathbf{E}_2 \times \mathbf{H}_1) \cdot d\mathbf{s},$$

(S1-4)

where the first and third step uses the expression of $\hat{\mathbf{D}}\hat{\mathbf{H}}$, while the second step utilizes one variant of the divergence theorem, i.e., $\iiint_V \mathbf{H}_2 \cdot \nabla \times \mathbf{E}_1 d^3\mathbf{r} - \iiint_V \mathbf{E}_1 \cdot \nabla \times \mathbf{H}_2 d^3\mathbf{r} = \iint_\Sigma (\mathbf{E}_1 \times \mathbf{H}_2) \cdot d\mathbf{s}$, where $\Sigma$ is the exterior surface enclosing V. And the fourth step is based on the relation $\hat{\mathbf{H}}\mathbf{\Psi}_2 = \omega_2 \mathbf{\Psi}_2 + \mathbf{S}_2$. Besides, the right-hand side of Eq. (S1-4) should be equal to the right-hand side of Eq. (A-1), leading to

$$(\omega_1 - \omega_2) \iiint_V \mathbf{\Psi}_1^T \cdot \hat{\mathbf{D}} \mathbf{\Psi}_2 d^3\mathbf{r} + \iiint_V (\mathbf{\Psi}_2^T \cdot \hat{\mathbf{D}} \mathbf{S}_1 - \mathbf{\Psi}_1^T \cdot \hat{\mathbf{D}} \mathbf{S}_2) d^3\mathbf{r} = i \iint_\Sigma (\mathbf{E}_1 \times \mathbf{H}_2 - \mathbf{E}_2 \times \mathbf{H}_1) \cdot d\mathbf{s}.$$

(S1-5)

Because $\hat{\mathbf{D}}$ is a diagonal matrix, there is $\iiint_V \mathbf{\Psi}_2^T \cdot \hat{\mathbf{D}} \mathbf{\Psi}_1 d^3\mathbf{r} = \iiint_V \mathbf{\Psi}_1^T \cdot \hat{\mathbf{D}} \mathbf{\Psi}_2 d^3\mathbf{r}$, which is also used in the derivation of Eq. (S1-5). Equation (S1-5) is just the unconjugated Lorentz reciprocity after introducing the auxiliary fields.

# Supplement 2    Poynting theorem

Here we derive the Poynting theorem after introducing the auxiliary fields. For the Maxwell's equations $\hat{\mathbf{H}}\boldsymbol{\Psi} = \omega\boldsymbol{\Psi} + \mathbf{S}$, applying $\iiint_V d^3\mathbf{r}\boldsymbol{\Psi}^\dagger\hat{\mathbf{A}}$ at both sides leads to

$$\iiint_V \boldsymbol{\Psi}^\dagger \cdot \hat{\mathbf{A}}\hat{\mathbf{H}}\boldsymbol{\Psi} d^3\mathbf{r} = \omega \iiint_V \boldsymbol{\Psi}^\dagger \cdot \hat{\mathbf{A}}\boldsymbol{\Psi} d^3\mathbf{r} + \iiint_V \boldsymbol{\Psi}^\dagger \cdot \hat{\mathbf{A}}\mathbf{S} d^3\mathbf{r} \quad \text{(S2-1)}$$

where $\boldsymbol{\Psi}^\dagger$ is is the conjugate transpose of $\boldsymbol{\Psi}$, $\omega$ is the frequency defined on the complex plane, and $\hat{\mathbf{A}}$ is defined as

$$\hat{\mathbf{A}} = \text{diag}\,[\mu_0, \epsilon_0\epsilon_{r\infty}, -\Omega^*/(A\epsilon_0\epsilon_{r\infty}), -\Omega/(A^*\epsilon_0\epsilon_{r\infty})] \quad \text{(S2-2)}$$

Expanding Eq. (S2-1) term by term gives

$$-2\text{Im}\,(\omega)W_e = P_{\text{abs}} + P_{\text{rad}} - P_{\text{inp}} \quad \text{(S2-3)}$$

Equation (S2-3) is just the Poynting theorem after introducing the auxiliary field, which means the conservation of energy, i.e., the energy decay rate $-2\text{Im}\,(\omega)W_e$ equals the subtraction of the total loss power (the summation of the material absorption power $P_{\text{abs}}$ and radiation power power $P_{\text{rad}}$) by the incident power $P_{\text{inp}}$. The expressions of $W_e$, $P_{\text{abs}}$, $P_{\text{rad}}$ and $P_{\text{inp}}$ are

$$W_e = \frac{1}{4}\iiint_V \left[\mu_0|\mathbf{H}|^2 + \epsilon_0\epsilon_{r\infty}|\mathbf{E}|^2 - \text{Re}\left(\frac{\Omega^*}{A\epsilon_0\epsilon_{r\infty}}\right)(|\mathbf{P}|_1^2 + |\mathbf{P}|_2^2)\right] d^3\mathbf{r}, \quad \text{(S2-4a)}$$

$$\begin{aligned}P_{\text{abs}} = \iiint_V &\left[\text{Im}\,(A\epsilon_0\epsilon_{\text{roo}})|\mathbf{E}|^2 + \frac{1}{2}\text{Im}\left(\frac{|\Omega|^2}{A\epsilon_0\epsilon_{\text{roo}}}\right)(|\mathbf{P}|_1^2 + |\mathbf{P}|_2^2)\right.\\ &\left.-\frac{1}{2}\text{Re}\,(\omega)\text{Im}\left(\frac{\Omega^*}{A\epsilon_0\epsilon_{\text{roo}}}\right)(|\mathbf{P}|_1^2 - |\mathbf{P}|_2^2)\right] d^3\mathbf{r},\end{aligned} \quad \text{(S2-4b)}$$

$$P_{\text{rad}} = \frac{1}{2}\iint_\Sigma \text{Re}\,(\mathbf{E} \times \mathbf{H}^*) \cdot d\mathbf{s}, \quad \text{(S2-4c)}$$

$$P_{\text{inp}} = -\frac{1}{2}\iiint_V \text{Im}\,(\boldsymbol{\Psi}^\dagger \cdot \hat{\mathbf{A}}\mathbf{S}) d^3\mathbf{r}. \quad \text{(S2-4d)}$$

Note that the total field is used in computing the $P_{\text{abs}}$ in Eq. (S2-4b). For dispersion model with multiple pairs of partial fractions, $W_e$ and $P_{\text{abs}}$ become

$$W_e = \frac{1}{4}\iiint_V \left[\mu_0|\mathbf{H}|^2 + \epsilon_0\epsilon_{r\infty}|\mathbf{E}|^2 - \sum_{j=1}^N \text{Re}\left(\frac{\Omega_j^*}{A_j\epsilon_0\epsilon_{r\infty}}\right)(|\mathbf{P}|_{1j}^2 + |\mathbf{P}|_{2j}^2)\right] d^3\mathbf{r}, \quad \text{(S2-5a)}$$

$$P_{\text{abs}} = \iiint_V \sum_{j=1}^{N} \left[ \text{Im}\left(A_j \epsilon_0 \epsilon_{r\infty}\right) |\mathbf{E}|^2 + \frac{1}{2} \text{Im}\left(\frac{|\Omega_j|^2}{A_j \epsilon_0 \epsilon_{r\infty}}\right) \left(|\mathbf{P}|_{1j}^2 + |\mathbf{P}|_{2j}^2\right) \right.$$
$$\left. - \frac{1}{2} \text{Re}(\omega) \text{Im}\left(\frac{\Omega_j^*}{A_j \epsilon_0 \epsilon_{r\infty}}\right) \left(|\mathbf{P}|_{1j}^2 - |\mathbf{P}|_{2j}^2\right) \right] d^3\mathbf{r}.$$
(S2-5b)

For a QNM with complex eigenfrequency $\tilde{\omega}$, because it is the solution to source-free Maxwell's equations, i.e., $\mathbf{S} = \mathbf{0}$, there is $P_{\text{inp}} = 0$, and Eq. (S2-3) becomes $-2\text{Im}(\tilde{\omega})W_e = P_{\text{abs}} + P_{\text{rad}}$. Thus the Q-factor of the QNM is defined as

$$Q = -\frac{\text{Re}(\tilde{\omega})}{2\text{Im}(\tilde{\omega})} = \text{Re}(\tilde{\omega})\frac{W_e}{P_{\text{abs}} + P_{\text{rad}}} = 2\pi \frac{\text{Energy stored}}{\text{Loss energy per period}}, \quad (S2\text{-}6)$$

which is consistent with that defined in classical electrodynamics[1]. Besides, Eq. (S2-6) is applicable in any area of the PML mapped space [2,3], and the total Q is a constant, but in different domain the ratio $P_{\text{abs}}/P_{\text{rad}}$ is usually different. So if we want to define $Q_{\text{abs}} = \text{Re}(\tilde{\omega})W_e/P_{\text{abs}}$ and $Q_{\text{abs}} = \text{Re}(\tilde{\omega})W_e/P_{\text{abs}}$ like those in the classical electrodynamics, it should be very cautious to choose the proper domain.

# Supplement 3   Equivalent Source of the scattered field

Equations (21) and (22) give the equivalent Source of the scattered field, or the background field $\mathbf{S}_{bg}$, here we will give its detailed derivation.

The total field $\mathbf{\Psi}_{total}$ satisfies the Maxwell's equations shown in Eq. (19), but the background field only fulfills the Maxwell's equations below when the resonator is absent (i.e., $\Delta\epsilon_r = 0$)

$$\nabla \times \mathbf{H}_{bg} = -i\omega\epsilon_0\epsilon_{rbg}\mathbf{E}_{bg} + \mathbf{J}_E, \qquad (S3\text{-}1)$$

$$\nabla \times \mathbf{E}_{bg} = i\omega\mu_0\mathbf{H}_{bg} + \mathbf{J}_M. \qquad (S3\text{-}2)$$

Like the definition of the augmented total field, we define the augmented background field $\mathbf{\Psi}_{bg} = [\mathbf{H}_{bg}, \mathbf{E}_{bg}, \mathbf{P}_{1bg}, \mathbf{P}_{2bg}]^T$, but $\mathbf{\Psi}_{bg}$ does not contain auxiliary field in the resonator domain, i.e., where $\mathbf{r} \in V_{res}$, there is $\mathbf{\Psi}_{bg} = [\mathbf{H}_{bg}, \mathbf{E}_{bg}, 0,0]^T$. Applying $\widehat{\mathbf{H}}$ (same with that defined in Eq. (8)) to $\mathbf{\Psi}_{bg}$ and transforming properly gives

$$\widehat{\mathbf{H}}\mathbf{\Psi}_{bg} = \omega\mathbf{\Psi}_{bg} + \begin{bmatrix} i\mu_0^{-1}\mathbf{J}_M \\ i(\epsilon_0\epsilon_{r\infty})^{-1}\mathbf{J}_E \\ 0 \\ 0 \end{bmatrix} - \begin{bmatrix} 0 \\ [\omega(\epsilon_{r\infty} - \epsilon_{rbg})/\epsilon_{r\infty} + (A - A^*)]\mathbf{E}_{bg} \\ -A\epsilon_0\epsilon_{r\infty}\mathbf{E}_{bg} \\ A^*\epsilon_0\epsilon_{ro\infty}\mathbf{E}_{bg} \end{bmatrix}. \quad (S3\text{-}3)$$

Note that the operation $\widehat{\mathbf{H}}$ in Eq. (S3-3) is for the case with resonator (i.e. $\Delta\epsilon_r \neq 0$), thus a third extra term appear compared with Eq. (1-19). From the definition of the scattering we have $\widehat{\mathbf{H}}\mathbf{\Psi}_{sca} = \widehat{\mathbf{H}}(\mathbf{\Psi}_{total} - \mathbf{\Psi}_{bg})$, and substituting Eqs. (1-19) and (S3-3) into it gives

$$\widehat{\mathbf{H}}\mathbf{\Psi}_{sca} = \omega\mathbf{\Psi}_{sca} + \begin{bmatrix} 0 \\ [\omega(\epsilon_{roo} - \epsilon_{rbg})/\epsilon_{roo} + (A - A^*)]\mathbf{E}_{bg} \\ -A\epsilon_0\epsilon_{roc}\mathbf{E}_{bg} \\ A^*\epsilon_0\epsilon_{roo}\mathbf{E}_{bg} \end{bmatrix}, \quad (S3\text{-}4)$$

$$\mathbf{S}_{bg} = \left[0, [\omega(\epsilon_{roo} - \epsilon_{rbg})/\epsilon_{roo} + (A - A^*)]\mathbf{E}_{bg}, -A\epsilon_0\epsilon_{roo}\mathbf{E}_{bg}, A^*\epsilon_0\epsilon_{ro\infty}\mathbf{E}_{bg}\right]^T, (S3\text{-}5)$$

where $\mathbf{S}_{bg}$ is just the equivalent source of the scattered field.

# Supplement 4   Term-by-term expansion of the extinction cross section

To evaluate the contribution of each eigenmode to the extinction cross section, we can further modify Eq. (32) into the form of a term-by-term expansion of QNMs.

Substituting Eq. (1-6) into Eq. (1-32) gives

$$P_{\text{ext}} = \frac{1}{2}\iiint_{V_{\text{res}}} \text{Im}\left\{\left[\omega\epsilon_0(\epsilon_{\text{roo}} - \epsilon_{\text{rbg}}^*)\right]\mathbf{E}_{\text{sca}}\cdot\mathbf{E}_{\text{bg}}^* + \omega(\mathbf{P}_{1,\text{sca}} + \mathbf{P}_{2,\text{sca}})\cdot\mathbf{E}_{\text{bg}}^*\right\}d^3\mathbf{r}.$$

(S4-1)

Substituting the eigenmodes expansion of the scattering field shown in Eq. (1-25) into Eq. (S4-1) gives

$$P_{\text{ext}} = \sum_{m=1}^{\infty}\frac{\omega}{2}\iiint_{V_{\text{res}}} \text{Im}\left\{\Lambda_m(\omega)\left[\epsilon_0(\epsilon_{\text{roo}} - \epsilon_{\text{tgg}}^*)\tilde{\mathbf{E}}_m + (\tilde{\mathbf{P}}_{1m} + \tilde{\mathbf{P}}_{2m})\right]\cdot\mathbf{E}_{\text{bg}}^*\right\}d^3\mathbf{r}$$

(S4-2)

For the eigenmodes, $\mathbf{E}_{\text{bg}} = 0, \tilde{\mathbf{P}}_{1m}, \tilde{\mathbf{P}}_{2m}$ and $\tilde{\mathbf{E}}_m$ satisfy the relation defined in Eq. (5). Considering the dispersion relation defined in Eq. (4) and the definition of the extinction cross section $\sigma_{\text{ext}} = P_{\text{ext}}/I_{\text{inc}}$, the term-by-term expansion of $\sigma_{\text{ext}}$ can be derived from Eq. (S4-2) as

$$\sigma_{\text{ext}} = \sum_{m=1}^{\infty}\frac{\omega}{2I_{\text{inc}}}\iiint_{V_{\text{res}}} \text{Im}\left\{\epsilon_0\left[(\epsilon_r(\tilde{\omega}_m) - \epsilon_{\text{rbg}}^*)\right]\Lambda_m(\omega)\tilde{\mathbf{E}}_m\cdot\mathbf{E}_{\text{bg}}^*\right\}d^3\mathbf{r}. \quad (S4-3)$$

Equation (S4-3) is also applicable to multiple partial fractions dispersion model cases (Eq. (1-33)). It is noteworthy that numerical experiments show that using Eq. (S4-3) to decompose $\sigma_{\text{ext}}$ would cause numerical instability, especially at high frequency. Therefore, we only use Eq. (S4-3) to evaluate the contribution of each eigenmode for ranking purpose, but still use Eq. (32) or Eq. (33) to reconstruct the total extinction cross section.

On the other hand, we can also study the extinction power shown in Eq. (S4-1) in the real frequency domain. As the augmented background field has no auxiliary field in $V_{\text{res}}$, i.e., when $\mathbf{r}\in V_{\text{res}}$, $\mathbf{\Psi}_{\text{bg}} = [\mathbf{H}_{\text{bg}}, \mathbf{E}_{\text{bg}}, 0, 0]^{\text{T}}$. Therefore, in $V_{\text{res}}$, $\mathbf{P}_{1,\text{sca}}, \mathbf{P}_{2,\text{sca}}$ and the total field $\mathbf{E}_{\text{total}} = \mathbf{E}_{\text{sca}} + \mathbf{E}_{\text{bg}}$ satisfy the relation defined in Eq. (5), and

substituting it into Eq. (S4-1) gives

$$P_{\text{ext}} = \frac{\omega}{2} \iiint_{V_{\text{res}}} \text{Im} \{\epsilon_0(\epsilon_{r\infty} - \epsilon_{\text{rbg}}^*)\mathbf{E}_{\text{sca}} \cdot \mathbf{E}_{\text{bg}}^* + \epsilon_0[\epsilon_r(\omega) - \epsilon_{r\infty}](\mathbf{E}_{\text{sca}} + \mathbf{E}_{\text{bg}}) \cdot \mathbf{E}_{\text{bg}}^*\} d^3\mathbf{r}. \quad \text{(S4-4)}$$

Substituting $\epsilon_0 \Delta\epsilon_r = \epsilon_0[\epsilon_r(\omega) - \epsilon_{\text{rbg}}]$ into Eq. (S4-4) and considering that $\epsilon_{r\infty}$ is real, there is

$\text{Im}\left[\epsilon_0(\epsilon_{r\infty} - \epsilon_{\text{rbg}}^*)\mathbf{E}_{\text{sca}} \cdot \mathbf{E}_{\text{bg}}^* + \epsilon_0(\epsilon_{r\infty} - \epsilon_{\text{rbg}})\mathbf{E}_{\text{sca}}^* \cdot \mathbf{E}_{\text{bg}}\right] = 0$ and $|\mathbf{E}_{\text{sca}} + \mathbf{E}_{\text{bg}}|^2 = (\mathbf{E}_{\text{sca}} + \mathbf{E}_{\text{bg}}) \cdot \mathbf{E}_{\text{bg}}^* + \mathbf{E}_{\text{sca}}^* \cdot \mathbf{E}_{\text{bg}} + |\mathbf{E}_{\text{sca}}|^2$, and eventually we can obtain that

$$\begin{aligned} P_{\text{ext}} &= \frac{\omega}{2} \iiint_{V_{\text{res}}} \text{Im}\left[\epsilon_0 \Delta\epsilon_r (\mathbf{E}_{\text{sca}} + \mathbf{E}_{\text{bg}}) \cdot \mathbf{E}_{\text{bg}}^*\right] d^3\mathbf{r} \\ &+ \frac{\omega}{2} \iiint_{V_{\text{res}}} \text{Im}(\epsilon_0 \epsilon_{\text{rbg}})\left[|\mathbf{E}_{\text{sca}} + \mathbf{E}_{\text{bg}}|^2 - |\mathbf{E}_{\text{sca}}|^2\right] d^3\mathbf{r}. \end{aligned} \quad \text{(S4-5)}$$

Equation (S4-5) is also consistent with that defined by Lalanne et. al. [1], but if we substitute the eigenmode expansion of $\mathbf{E}_{\text{sca}}$ (Eq. (1-25)) directly into Eq. (S4-5) to compute the extinction power, the problem of numerical instability will also appear.